\documentclass{article}

\usepackage{arxiv}

\usepackage[utf8]{inputenc} 
\usepackage[T1]{fontenc}    
\usepackage{hyperref}       
\usepackage{url}            
\usepackage{booktabs}       
\usepackage{amsfonts}       
\usepackage[numbers]{natbib}
\usepackage{amsmath, enumitem}
\usepackage{upgreek}
\usepackage{graphicx}
\graphicspath{{Figures/}} 
\usepackage{setspace}
\usepackage{multirow}
\usepackage{fixltx2e}

\DeclareMathOperator*{\argmax}{arg\,max}

\title{Seismic tomography using variational inference methods}

\author{
  Xin Zhang \\
  School of Geosciences \\
  University of Edinburgh\\
  Edinburgh, United Kingdom \\
  \texttt{x.zhang2@ed.ac.uk} \\
   \And
 Andrew Curtis \\
  School of Geosciences \\
  University of Edinburgh\\
  Edinburgh, Unite Kingdom \\
  \texttt{andrew.curtis@ed.ac.uk} \\
}

\begin{document}
\maketitle

\begin{abstract}
Seismic tomography is a methodology to image the interior of solid or fluid media, and is often used to map properties in the subsurface of the Earth. In order to better interpret the resulting images it is important to assess imaging uncertainties. Since tomography is significantly nonlinear, Monte Carlo sampling methods are often used for this purpose, but they are generally computationally intractable for large datasets and high-dimensional parameter spaces. To extend uncertainty analysis to larger systems we use variational inference methods to conduct seismic tomography. In contrast to Monte Carlo sampling, variational methods solve the Bayesian inference problem as an optimization problem, yet still provide probabilistic results. In this study, we applied two variational methods, automatic differential variational inference (ADVI) and Stein variational gradient descent (SVGD), to 2D seismic tomography problems using both synthetic and real data and we compare the results to those from two different Monte Carlo sampling methods. The results show that variational inference methods can produce accurate approximations to the results of Monte Carlo sampling methods at significantly lower computational cost, provided that gradients of parameters with respect to data can be calculated efficiently. We expect that the methods can be applied fruitfully to many other types of geophysical inverse problems.
\end{abstract}

\section{Introduction}
In a variety of geoscientific applications, scientists need to obtain maps of subsurface properties in order to understand heterogeneity and processes taking place within the Earth. Seismic tomography is a method that is widely used to generate those maps. The maps of interest are usually parameterised in some way, and data are recorded that can be used to constrain the parameters. Tomography is therefore a parameter estimation problem, given the data and a physical relationship between data and parameters; since the physical relationships usually predict data given parameter values but not the reverse, seismic tomography involves solving an inverse problem \cite{curtis2002probing}. 

Tomographic problems can be solved using either the full, known physical relationships, or by first creating approximate, linearised physics. In the linearised case, one usually seeks an optimal solution by minimizing the misfits between predicted data and observed data by iteratively linearising the physics around the current best model and solving the linear system to update that model estimate. Since most tomography problems are under-determined, some form of regularization must be introduced to solve the system \cite{aki1976determination, dziewonski1987global, iyer1993seismic, tarantola2005inverse}. However, regularization is usually chosen using ad hoc criteria which introduces poorly understood biases in the results; thus, valuable information can be concealed by regularization \cite{zhdanov2002geophysical}. Moreover, in nonlinear problems it is almost always impossible to estimate accurate uncertainties in results using linearised methods. Therefore, partially or fully nonlinear tomographic methods have been introduced to geophysics which require no linearisation and which provide accurate estimates of uncertainty using a Bayesian probabilistic formulation of the parameter estimation problem. These include Monte Carlo methods \cite{mosegaard1995monte, sambridge1999geophysical, malinverno2000monte, malinverno2002parsimonious,malinverno2004expanded, bodin2009seismic, galetti2015uncertainty, galetti2017transdimensional, zhang20183} and methods based on neural networks \cite{roth1994neural, devilee1999efficient, meier2007aglobal, meier2007bglobal, shahraeeni2011fast, shahraeeni2012fast, kaufl2013framework, kaufl2015robust, earp2019probabilistic}. 

Bayesian methods use Bayes' theorem to update a \textit{prior} probability distribution function (\textit{pdf} -- either a conditional density function or a discrete set of probabilities) with new information from data. The prior pdf describes information available about the parameters of interest prior to the inversion. Bayes' theorem combines the prior pdf with information derived from the data to produce the total state of information about the parameters post inversion, described by a so-called \textit{posterior} pdf -- this process is referred to as Bayesian inference. Thus, in our case Bayesian inference is used to solve the tomographic inverse problem. 

Monte Carlo methods generate a set (or chain) of samples from the posterior pdf describing the probability distribution of the model given the observed data; thereafter these samples can be used to estimate useful information about that pdf (mean, standard deviation, etc.). The methods are quite general from a theoretical point of view so that in principal they can be applied to any tomographic problems. They have been extended to trans-dimensional inversion using the reversible jump Markov chain Monte Carlo (rj-McMC) algorithm \cite{green1995reversible}, in which the number and meaning of parameters (hence the dimensionality of parameter space) can vary in the inversion. Consequently the parameterization itself can be simplified by adapting to the data which improves results on otherwise high-dimensional problems \cite{malinverno2000monte, bodin2009seismic, bodin2012transdimensional,ray2013robust, young2013transdimensional, galetti2015uncertainty, galetti2017transdimensional, hawkins2015geophysical, piana2015local, burdick2017velocity,  galetti2018transdimensional, zhang20183, zhang2019fully}. Although many applications have been conducted using McMC sampling methods \cite[previous references;][]{shen2012joint, shen20133, zulfakriza2014upper, zheng2017transdimensional, crowder2019transdimensional}, they mainly address 1D or 2D tomography problems due to the high computational expense of Monte Carlo methods. Some studies used McMC methods for fully 3D tomography using body wave travel time data \cite{hawkins2015geophysical, piana2015local, burdick2017velocity} and surface wave dispersion \cite{zhang20183, zhang2019fully}, but the methods demand enormous computational resources. Even in the 1D or 2D case, McMC methods cannot easily be applied to large datasets which are generally expensive to forward model given a set of parameter values. Moreover, McMC methods tend to be inefficient at exploring complex, multi-modal probability distributions \cite{sivia1996data, karlin2014first}, which appear to be common in seismic tomography problems.  

Neural network based methods offer an efficient alternative for certain classes of tomography problems that will be solved many times with new data of the same type. An initial set of Monte Carlo samples is taken from the prior probability distribution over parameter space, and data are computationally forward modelled for each parameter vector. Neural networks are flexible mappings that can be regressed (trained) to emulate the mapping from data to parameter space by fitting the set of examples of that mapping generated using Monte Carlo \cite{bishop2006pattern}. The trained network then interpolates the inverse mapping between the examples, and can be applied efficiently to any new, measured data to estimate corresponding parameter values. The first geophysical application of neural network tomography was \citet{roth1994neural}, but that application did not estimate uncertainties. Forms of networks that estimate tomographic uncertainties were introduced by \citet{devilee1999efficient} and \citet{meier2007aglobal, meier2007bglobal} and have been applied to surface and body wave tomography in 1D and 2D problems \cite{meier2007aglobal, meier2007bglobal, earp2019probabilistic}. Nevertheless, neural networks still suffer from the computational cost of generating the initial set of training examples. That set may have to include many more samples than are required for standard Bayesian MC, because the training set must span the prior pdf whereas standard applications of MC tomography sample the posterior pdf which is usually more tightly constrained. Neural networks have the advantage that the training samples need only be calculated once for any number of data sets whereas MC inversion must perform sampling for every new data set. However, in high dimensional problems the cost of sampling may be prohibitive for both MC and NN based methods due to the curse of dimensionality \cite[the exponential increase in the hypervolume of parameter space as the number of parameters increases -- ][]{curtis2001prior}.

Variational inference provides a different way to solve a Bayesian inference problem: within a predefined family of probability distributions, one seeks an optimal approximation to a target distribution which in this case is the Bayesian posterior pdf. This is achieved by minimizing the Kullback-Leibler (KL) divergence \cite{kullback1951information} -- a measure of the difference between the approximate and target pdfs \cite{bishop2006pattern, blei2017variational}. Since the method casts the inference problem into an optimization problem, it can be computationally more efficient than either MC sampling or neural network methods, and provides better scaling to higher dimensional problems. Moreover, it can be used to take advantage of methods such as stochastic optimization \cite{robbins1951stochastic, kubrusly1973stochastic} and distributed optimization by dividing large datasets into random minibatches -- methods which are difficult to apply for McMC methods since they may break the reversibility property of Markov chains which is required by most McMC methods. 

In variational inference, the complexity of the approximating family of pdfs determines the complexity of the optimization. A complex variational family is generally more difficult to optimize than a simple family. Therefore, many applications are performed using simple mean-field approximation families \cite{bishop2006pattern, blei2017variational} and structured families \cite{saul1996exploiting, hoffman2015structured}. For example, in Geophysics the method has been used to invert for the spatial distribution of geological facies given seismic data using a mean-field approximation \cite{nawaz2018variational, nawaz2019rapid}. 

Even using those simple families, applications of variational inference methods usually involve tedious derivations and bespoke implementations for each type of problem which restricts their applicability \cite{bishop2006pattern, blei2017variational, nawaz2018variational, nawaz2019rapid}. The simplicity of those families also affects the quality of the approximation to complex distributions. To make variational methods easier to use, "black box" variational inference methods have been proposed \cite{kingma2013auto, ranganath2014black, ranganath2016hierarchical}. Based on these ideas, \citet{kucukelbir2017automatic} proposed an automatic variational inference method which can easily be applied to many Bayesian inference problems. Another set of methods has been proposed based on probability transformations \cite{rezende2015variational, tran2015variational, liu2016stein, marzouk2016introduction}; these methods optimise a series of invertible transforms to approximate the target probability and in this case it is possible to approximate arbitrary probability distributions. 

We apply automatic differential variational inference \cite[ADVI --][]{kucukelbir2017automatic} and Stein variational gradient descent \cite[SVGD --][]{liu2016stein} to a 2D seismic tomography problem. In the following we first describe the basic idea of variational inference, and then the ADVI and SVGD methods. In section 3 we apply the two methods to a simple 2D synthetic seismic tomography example and compare their results with both fixed-dimensional McMC and rj-McMC. In section 4 we apply the two methods to real data from Grane field, North Sea, to study the phase velocity map at 0.9 s and compare the results to those found using rj-McMC. We thus demonstrate that variation inference methods can provide efficient alternatives to McMC methods while still producing reasonably accurate approximations to Bayesian posterior pdfs. Our aim is to introduce variational inference methods to the geoscientific community and to encourage more research on this topic.

\section{Methods}
\subsection{Variational inference}
Bayesian inference involves calculating or characterising a posterior probability density function $p(\mathbf{m}|\mathbf{d}_{obs})$ of model parameters $\mathbf{m}$ given the observed data $\mathbf{d}_{obs}$. According to Bayes' theorem,
\begin{equation}
    p(\mathbf{m}|\mathbf{d}_{obs}) = \frac{p(\mathbf{d}_{obs}|\mathbf{m})p(\mathbf{m})}{p(\mathbf{d}_{obs})}
\label{eq:Bayes}
\end{equation}
where $p(\mathbf{d}_{obs}|\mathbf{m})$ is called the \textit{likelihood} which is the probability of observing data $\mathbf{d}_{obs}$ if model $\mathbf{m}$ was true, $p(\mathbf{m})$ is the \textit{prior} which describes information that is independent of the data, and $p(\mathbf{d}_{obs})$ is a normalization factor called the \textit{evidence} which is constant for a fixed model parameterization. The likelihood is usually assumed to follow a Gaussian probability density function around the data predicted synthetically from model $\mathbf{m}$ (using the known physical relationships), as this is assumed to be a reasonable approximation to the pdf of uncertainties or errors in the measured data.

Variational inference approximates the above pdf $p(\mathbf{m}|\mathbf{d}_{obs})$ using optimization. First a family (set) of known distributions $\mathcal{Q}=\{q(\mathbf{m})\}$ is defined. The method then seeks the best approximation to $p(\mathbf{m}|\mathbf{d}_{obs})$ within that family by minimizing the KL-divergence:
\begin{equation}
\mathrm{KL}[q(\mathbf{m})||p(\mathbf{m}|\mathbf{d}_{obs})] = \mathrm{E}_{q}[\mathrm{log}q(\mathbf{m})] - \mathrm{E}_{q}[\mathrm{log}p(\mathbf{m}|\mathbf{d}_{obs})]
\label{eq:KL}
\end{equation}
where the expectation is taken with respect to distribution $q(\mathbf{m})$. It can be shown that $\mathrm{KL}[q||p]\geq0$ and has zero value if and only if $q(\mathbf{m})$ equals $p(\mathbf{m}|\mathbf{d}_{obs})$ \cite{kullback1951information}. Distribution $q^{*}(\mathbf{m})$ that minimizes the KL-divergence is therefore the best approximation to $p(\mathbf{m}|\mathbf{d}_{obs})$ within the family $\mathcal{Q}$. 

Combining equations (\ref{eq:Bayes}) and (\ref{eq:KL}), the KL-divergence becomes:
\begin{equation}
\mathrm{KL}[q(\mathbf{m})||p(\mathbf{m}|\mathbf{d}_{obs})] = \mathrm{E}_{q}[\mathrm{log}q(\mathbf{m})] - \mathrm{E}_{q}[\mathrm{log}p(\mathbf{m},\mathbf{d}_{obs})] + \mathrm{log}p(\mathbf{d}_{obs})
\label{eq:KL2}
\end{equation}
The evidence term $\mathrm{log}p(\mathbf{d}_{obs})$ generally cannot be calculated since it involves the evaluation of a high dimensional integral which takes exponential time. Instead we calculate the evidence lower bound (ELBO) which is equivalent to the KL-divergence up to an unknown constant, and is obtained by rearranging equation (\ref{eq:KL2}) and using the fact that $\mathrm{KL}[q||p]\geq0$:
\begin{equation}
\begin{aligned}
\mathrm{ELBO}[q] & =  \mathrm{E}_{q}[\mathrm{log}p(\mathbf{m},\mathbf{d}_{obs})] - \mathrm{E}_{q}[\mathrm{log}q(\mathbf{m})] \\
 & = \mathrm{log}p(\mathbf{d}_{obs}) - \mathrm{KL}[q(\mathbf{m})||p(\mathbf{m}|\mathbf{d}_{obs})] 
\end{aligned}
\label{eq:ELBO}
\end{equation}
Thus minimizing the KL-divergence is equivalent to maximizing the ELBO.

In variational inference, the choice of the variational family is important because the flexibility of the variational family determines the power of the approximation. However, it is usually more difficult to optimize equation (\ref{eq:ELBO}) over a complex family than a simple family. Therefore, many applications are performed using the \textit{mean-field} variational family, which means that the parameters $\mathbf{m}$ are treated as being mutually independent \cite{bishop2006pattern, blei2017variational}. However, even under that simplifying assumption, traditional variational methods require tedious model-specific derivations and implementations, which restricts their applicability to those problems for which derivations have been performed \cite[e.g.,][]{nawaz2018variational, nawaz2019rapid}. We therefore introduce two more general variational methods: the automatic differential variational inference (ADVI) and the Stein variational gradient descent (SVGD), which can both be applied to general inverse problems.
 
\subsection{Automatic differential variational inference (ADVI)}

\begin{figure}
\includegraphics[width=1.\linewidth]{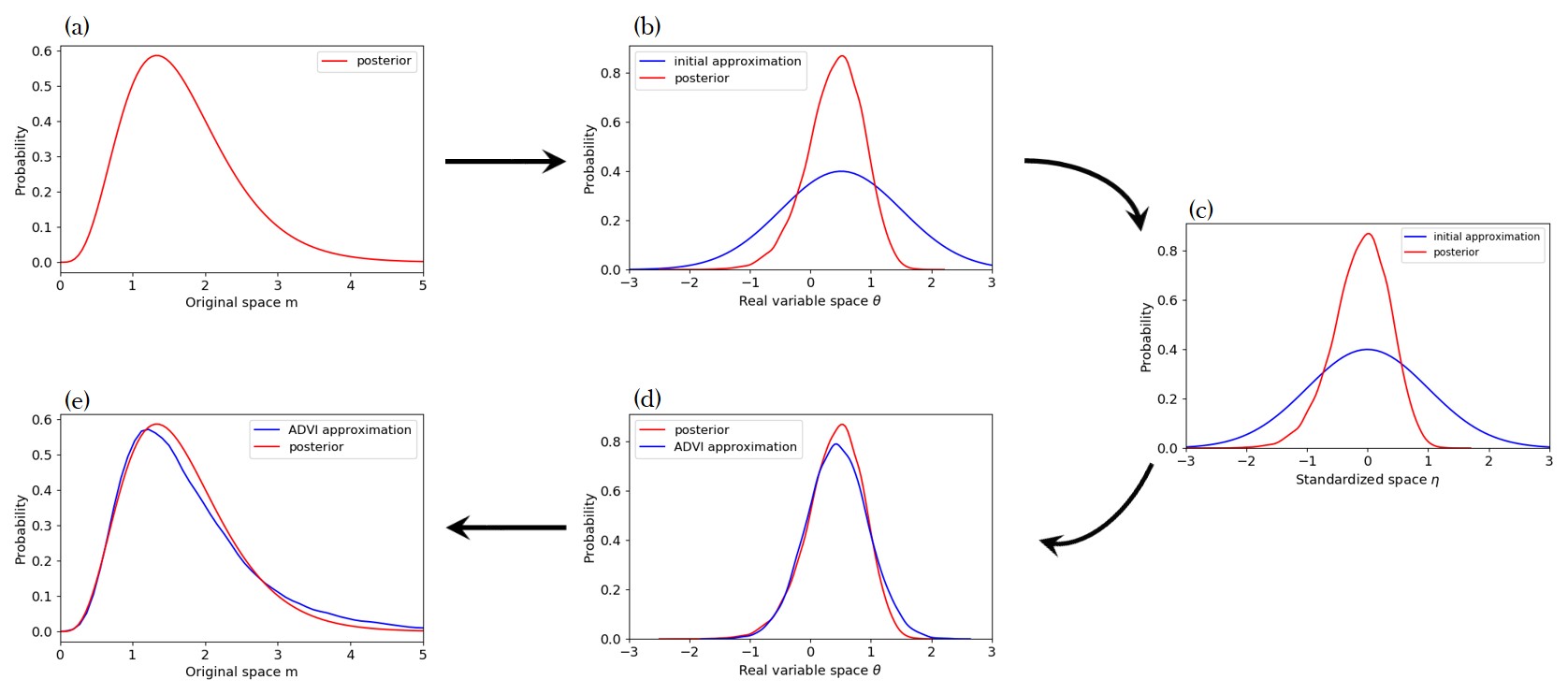}
\caption{An illustration of the workflow of ADVI. \textbf{(a)} An example of a posterior pdf in the original positive half space of parameters $\mathbf{m}$. \textbf{(b)} The posterior pdf in the transformed real variable space $\boldsymbol{\theta}$ (red) and an initial Gaussian approximation (blue). \textbf{(c)} The posterior pdf (red) and the standard Gaussian distribution (blue) in standardized variable $\boldsymbol{\eta}$; gradients with respect to variational parameters are calculated in this space. \textbf{(d)} and \textbf{(e)} show the posterior pdf (red) and the approximation obtained using ADVI (blue) in the unconstrained real variable space and the original space, respectively.}
\label{fig:advi}
\end{figure}

\citet{kucukelbir2017automatic} proposed a general variational method called automatic differential variational inference (ADVI) based on a Gaussian variational family. In ADVI, a model with constrained parameters is first transformed to a model with unconstrained real-valued variables. For example, the velocity model $\mathbf{m}$ that usually has hard bound constraints (such as velocity being greater than zero) can be transformed to an unconstrained model $\boldsymbol{\theta}=T(\mathbf{m})$, where $T$ is an invertible and differentiable function (Figure \ref{fig:advi}a and b). The joint probability $p(\mathbf{m},\mathbf{d}_{obs})$ then becomes: 
\begin{equation}
 p(\boldsymbol{\theta},\mathbf{d}_{obs}) = p(\mathbf{m},\mathbf{d}_{obs})|det\mathbf{J}_{T^{-1}}(\boldsymbol{\theta})|
 \label{eq:prob_transforma}
\end{equation}
where $\mathbf{J}_{T^{-1}}(\boldsymbol{\theta})$ is the Jacobian matrix of the inverse of $T$ which accounts for the volume change of the transform. This transform makes the choice of variational approximations independent of the original model since transformed variables lie in the common unconstrained space of real numbers.

In ADVI, we choose a Gaussian variational family (e.g., blue line in Figure \ref{fig:advi}b),
\begin{equation}
q(\boldsymbol{\theta};\boldsymbol{\phi})=\mathcal{N}(\boldsymbol{\theta}|\boldsymbol{\mu},\boldsymbol{\Sigma})=\mathcal{N}(\boldsymbol{\theta}|\boldsymbol{\mu},\mathbf{L}\mathbf{L}^{T})
\label{eq:normal}
\end{equation}
where $\boldsymbol{\phi}$ represents variational parameters $\boldsymbol{\mu}$ and $\boldsymbol{\Sigma}$, $\boldsymbol{\mu}$ is the mean vector and $\boldsymbol{\Sigma}$ is the covariance matrix. As in \citet{kucukelbir2017automatic} we use a Cholesky factorization $\boldsymbol{\Sigma}=\mathbf{L}\mathbf{L}^{T}$ where $\mathbf{L}$ is a lower-triangular matrix, to re-parameterize the covariance matrix to ensure that it is positive semidefinite. If $\boldsymbol{\Sigma}$ is a diagonal matrix, $q$ reduces to a mean-field approximation in which the variables are mutually independent; in order to include spatial correlations in the velocity model we use a full-rank covariance matrix, noting that this incurs a computational cost since it increases the number of variational parameters.

In the transformed space, the variational problem is solved by maximizing the ELBO, written as $\mathcal{L}$, with respect to variational parameters $\boldsymbol{\phi}$:
\begin{equation}
\begin{aligned}
 \boldsymbol{\phi}^{*} &= \argmax_{\boldsymbol{\phi}} \mathcal{L}[q(\boldsymbol{\theta};\boldsymbol{\phi})]\\
 &= \argmax_{\boldsymbol{\phi}} \mathrm{E}_{q}\left[\mathrm{log} p( T^{-1}(\boldsymbol{\theta}),\mathbf{d}_{obs} )+\mathrm{log}|det\mathbf{J}_{T^{-1}}(\boldsymbol{\theta})| \right] - \mathrm{E}_{q} \left[ \mathrm{log}q(\boldsymbol{\theta}) \right] 
 \end{aligned}
 \label{eq:argmaxELBO}
\end{equation}
This is an optimization problem in an unconstrained space and can be solved using gradient ascent methods without worrying about any constrains on the original variables. 

However, the gradients of variational parameters are not easy to calculate since the ELBO involves expectations in a high dimensional space. We therefore transform the Gaussian distribution $q(\boldsymbol{\theta};\boldsymbol{\phi})$ into a standard Gaussian $\mathcal{N}(\boldsymbol{\eta}|\boldsymbol{0},\mathbf{I})$ (Figure \ref{fig:advi}c), by $\boldsymbol{\eta}=R_{\boldsymbol{\phi}}(\boldsymbol{\theta})=\mathbf{L}^{-1}(\boldsymbol{\theta}-\boldsymbol{\mu})$, thereafter the variational problem becomes:
\begin{equation}
\begin{aligned}
 \boldsymbol{\phi}^{*} &= \argmax_{ \boldsymbol{\phi} } \mathcal{L} [q(\boldsymbol{\theta};\boldsymbol{\phi})] \\
 &= \argmax_{\boldsymbol{\phi}} \mathrm{E}_{ \mathcal{N} (\boldsymbol{\eta} | \boldsymbol{0},\mathbf{I} ) } \left[ \mathrm{log} p\Big( T^{-1} \left( R_{\boldsymbol{\phi}}^{-1} (\boldsymbol{\eta}) \right), \mathbf{d}_{obs} \Big) + \mathrm{log} |det\mathbf{J}_{T^{-1}} \left(R_{\boldsymbol{\phi}}^{-1} (\boldsymbol{\eta} )\right) | \right] - \mathrm{E}_{q} \left[ \mathrm{log} q(\boldsymbol{\theta}) \right] 
 \end{aligned}
 \label{eq:argmaxELBO2}
\end{equation} 
where the first expectation is taken with respect to a standard Gaussian distribution $\mathcal{N}(\boldsymbol{\eta}|\boldsymbol{0},\mathbf{I})$. There is no Jacobian term related to this transform since the determinant of the Jacobian is equal to one \cite{kucukelbir2017automatic}. The second expectation $-\mathrm{E}_{q}[\mathrm{log}q(\boldsymbol{\theta})]$ is not transformed since it has a simple analytic form as does its gradient \cite{kucukelbir2017automatic} -- see Appendix A.

Since the distribution with respect to which the expectation is taken now does not depend on variational parameters, the gradient with respect to variational parameters can be calculated by exchanging the expectation and derivative according to the dominated convergence theorem \cite{ccinlar2011probability} and by applying the chain rule -- see Appendix B:
\begin{equation}
 \nabla_{\boldsymbol{\mu}}\mathcal{L} = \mathrm{E}_{\mathcal{N}(\boldsymbol{\eta}|\boldsymbol{0},\mathbf{I})}\left[\nabla_{\mathbf{m}}\mathrm{log}p(\mathbf{m},\mathbf{d}_{obs})\nabla_{\boldsymbol{\theta}}T^{-1}(\boldsymbol{\theta})+\nabla_{\boldsymbol{\theta}}\mathrm{log}|det\mathbf{J}_{T^{-1}}(\boldsymbol{\theta})| \right]
 \label{eq:gradient_mu}
\end{equation}
The gradient with respect to $\mathbf{L}$ can be obtained similarly,
\begin{equation}
 \nabla_{\mathbf{L}}\mathcal{L} = \mathrm{E}_{\mathcal{N}(\boldsymbol{\eta}|\boldsymbol{0},\mathbf{I})}\left[ \nabla_{\mathbf{m}}\mathrm{log} p(\mathbf{m},\mathbf{d}_{obs})\nabla_{\boldsymbol{\theta}}T^{-1}(\boldsymbol{\theta})+\nabla_{\boldsymbol{\theta}}\mathrm{log}|det\mathbf{J}_{T^{-1}}(\boldsymbol{\theta})|\boldsymbol{\eta}^{T} \right] + (\mathbf{L}^{-1})^{T}
 \label{eq:gradient_sigma}
\end{equation}
where the expectation is computed with respect to a standard Gaussian distribution, which can be estimated by Monte Carlo (MC) integration. MC integration provides a noisy, unbiased estimation of the expectation and its accuracy increases with the number of samples. Nevertheless, it has been shown that in practice a low number or even a single sample can be sufficient at each iteration since the mean is taken with respect to the standard Gaussian distribution \cite[see discussions and experiments in][]{kucukelbir2017automatic}. For distributions $p(\mathbf{m},\mathbf{d}_{obs})$ for which the gradients have analytic forms, the whole process of computing gradients can be automated \cite{kucukelbir2017automatic}, hence the name "automatic differential". We can then use a gradient ascent method to update the variational parameters and obtain an approximation to the pdf $p(\mathbf{m}|\mathbf{d}_{obs})$ (e.g. Figure \ref{fig:advi}d). 

Note that although the method is based on Gaussian variational approximations, the actual shape of the approximation to the posterior $p(\mathbf{m}|\mathbf{d}_{obs})$ over the original parameters $\mathbf{m}$ is determined by the transform $T$ (Figure \ref{fig:advi}e). It is difficult to determine an optimal transform since that is related to the properties of the unknown posterior \cite{kucukelbir2017automatic}. In this study we use a commonly-used invertible logarithmic transform \cite{stan2016stan},
\begin{equation}
\begin{aligned}
 \theta_{i} &= T(m_{i}) = \mathrm{log}(m_{i}-a_{i}) - \mathrm{log}(b_{i}-m_{i}) \\
 m_{i} &= T^{-1}(\theta_{i}) = a_{i} + \frac{(b_{i}-a_{i})}{1+exp(-\theta_{i})}
\end{aligned}
 \label{eq:transform}
\end{equation} 
where $m_{i}$ represents each original constrained parameter, $\theta_{i}$ is the transformed unconstrained variable, $a_{i}$ is the original lower bound and $b_{i}$ the upper bound on $m_{i}$. Therefore the quality of the ADVI approximation is limited by the Gaussian approximation in the unconstrained space and by the specific transform $T$ in equation (\ref{eq:transform}).

\begin{figure}
\includegraphics[width=.9\linewidth]{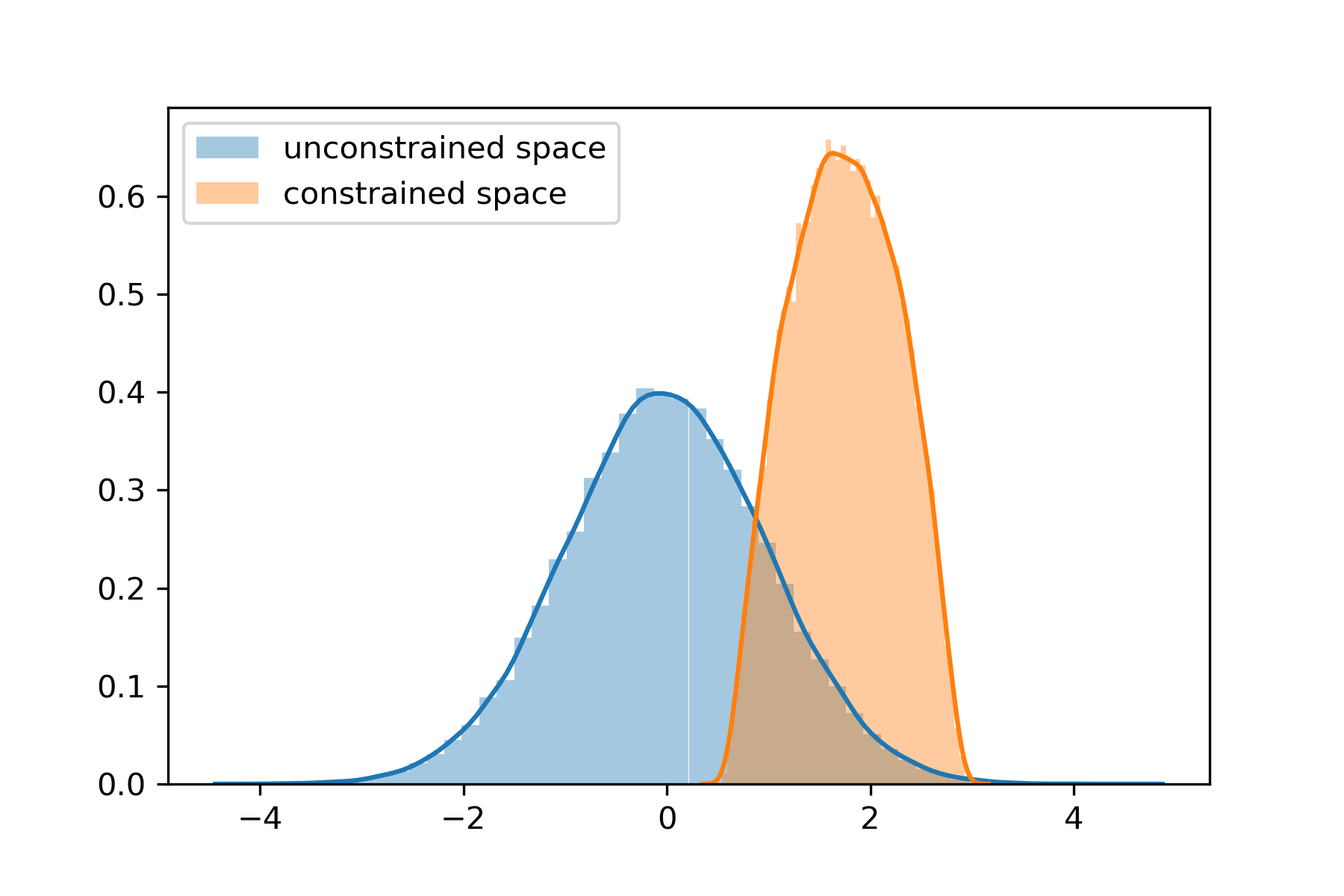}
\caption{An illustration of the transform in equation (\ref{eq:transform}). The original variable is in a constrained space between 0.5 and 3.0. The blue area shows a standard Gaussian distribution in the transformed unconstrained space and the orange area shows the associated probability distribution in the original space. The probability distributions are estimated using Monte Carlo samples.}
\label{fig:transform}
\end{figure}

To illustrate the effects of the transform in equation (\ref{eq:transform}), we show an example in Figure \ref{fig:transform}. The original variable lies in a constrained space between 0.5 and 3.0 (a typical phase velocity range of seismic surface waves). The space is transformed to an unconstrained space using equation (\ref{eq:transform}). If, as in ADVI we assume a standard Gaussian distribution in the transformed space (blue area in Figure \ref{fig:transform}), the associated probability distribution in the original space is shown in orange in Figure \ref{fig:transform}. The actual shape of the distribution in the original space is not Gaussian but is determined by the transform $T$ in equation (\ref{eq:transform}). However, under this choice of $T$ it is likely that the probability distribution in the original space is still unimodal. We thus see that ADVI provides a unimodal approximation of the target posterior pdf around a local optimal parameter estimate. This suggests that the method will not be effective for multimodal distributions, and the estimated probability distribution depends on the initial value of $\mu$ and $\Sigma$ \cite{kucukelbir2017automatic}. However, since the maximum a posteriori probability (MAP) estimate has been shown to be effective for parameter estimation in practice, the ADVI method could still be used to provide a good approximation of the distribution around a MAP estimate.  

\subsection{Stein variational gradient descent (SVGD)}

In practice most applications of variational inference use simple families of posterior approximations such as a Gaussian approximation \cite{kucukelbir2017automatic}, mean-field approximations \cite{blei2017variational, nawaz2018variational, nawaz2019rapid} or other simple structured families \cite{saul1996exploiting, hoffman2015structured}. These simple choices significantly restrict the quality of derived posterior approximations. In order to employ a broader family of variational approximations, variational methods based on invertible transforms have been proposed \cite{rezende2015variational, tran2015variational, marzouk2016introduction}. In these methods instead of choosing specific forms for variational approximations, a series of invertible transforms are applied to an initial distribution, and these transforms are optimized by minimizing the KL-divergence. This provides a way to approximate arbitrary posterior distributions since a pdf can be transformed to any other pdf as long as the probability measures are absolutely continuous.

Stein variational gradient descent (SVGD) is one such algorithm based on an incremental transform \cite{liu2016stein}. In SVGD, a smooth transform $T(\mathbf{m}) = \mathbf{m} + \epsilon \boldsymbol{\phi} (\mathbf{m})$ is used, where $\mathbf{m}=[m_{1}, ..., m_{d}]$ and $m_{i}$ is the $i^{th}$ parameter, and $\boldsymbol{\phi}(\mathbf{m}) = [\phi_{1},...,\phi_{d}]$ is a smooth vector function that describes the perturbation direction and where $\epsilon$ is the magnitude of the perturbation. It can be shown that when $\epsilon$ is sufficiently small, the transform is invertible since the Jacobian of the transform is close to an identity matrix \cite{liu2016stein}. Say $q_{T}(\mathbf{m})$ is the transformed probability distribution of the initial distribution $q(\mathbf{m})$. Then the gradient of KL-divergence with respect to $\epsilon$ can be computed as (see Appendix C):
\begin{equation}
\nabla_{\epsilon} \mathrm{KL}[q_{T}||p] \, |_{\epsilon=0} = - \mathrm{E}_{q} 
\left[ trace \left( \mathcal{A}_{p} \boldsymbol{\phi} (\mathbf{m}) \right) \right] 
\label{eq:stein_gradient}
\end{equation} 
where $\mathcal{A}_{p}$ is the Stein operator such that $\mathcal{A}_{p} \boldsymbol{\phi}(\mathbf{m}) = \nabla_{\mathbf{m}} \mathrm{log} p(\mathbf{m}) \boldsymbol{\phi} (\mathbf{m})^{T} + \nabla_{ \mathbf{m} } \boldsymbol{\phi} ( \mathbf{m} )$. This suggests that maximizing the right-hand expectation with respect to $q(\mathbf{m})$ gives the steepest descent of the KL-divergence, and consequently the KL-divergence can be minimized iteratively.

It can be shown that the negative gradient of the KL-divergence in equation (\ref{eq:stein_gradient}) can be maximized by using the kernelized Stein discrepancy \cite{liu2016kernelized}. For two continuous probability densities $p$ and $q$, the \textit{Stein discrepancy} for a function $\boldsymbol{\phi}$ in a function set $\mathcal{F}$ is defined as:
\begin{equation}
 S[q,p] = \argmax_{ \boldsymbol{\phi} \in \mathcal{F}} \{ \left[ \mathrm{E}_{q} trace \left(\mathcal{A}_{p} \boldsymbol{\phi}(\mathbf{m}) \right) \right] ^{2} \}
 \label{eq:stein_discrepancy}
\end{equation}
The Stein discrepancy provides another way to quantify the difference between two distribution densities \cite{stein1972bound, gorham2015measuring}. However the Stein discrepancy is not easy to compute for general $\mathcal{F}$. Therefore, \citet{liu2016kernelized} proposed a kernelized Stein discrepancy by maximizing equation (\ref{eq:stein_discrepancy}) in the unit ball of a reproducing kernel Hilbert space (RKHS) as follows. 

A Hilbert space is a space $\mathcal{H}$ on which an inner product $<,>_{\mathcal{H}}$ is defined. A function is called a \textit{kernel} if there exists a real Hilbert space and a function $\varphi$ such that $k(x,y)=<\varphi(x),\varphi(y)>_{\mathcal{H}}$ \cite{gretton2013introduction}. A kernel is said to be positive-definite if the matrix defined by $K_{ij}=k(x_{i},x_{j})$ is positive definite. Assuming a positive definite kernel $k(\mathbf{m},\mathbf{m'})$ on $\mathcal{M} \times \mathcal{M}$, its reproducing kernel Hilbert space $\mathcal{H}$ is defined by the closure of the linear span $ \{f: f(\mathbf{m}) = \sum_{i=1}^{n} a_{i}k(\mathbf{m},\mathbf{m}^{i}), a_{i} \in \mathcal{R}, n \in \mathcal{N}, \mathbf{m}^{i} \in \mathcal{M} \} $ with inner products $\langle f, g \rangle _{\mathcal{H}} = \sum_{ij} a_{i} b_{j} k(\mathbf{m}^{i},\mathbf{m}^{j})$ for $g(\mathbf{m})=\sum_{i} b_{i}k(\mathbf{m},\mathbf{m}^{i})$. The RKHS has an important reproducing property, that is, $f(x)=<f(x'),k(x',x)>_{\mathcal{H}}$, such that the evaluation of a function $f$ at $x$ can be represented as an inner product in the Hilbert space. In a RKHS, the kernelized Stein discrepancy can be defined as \cite{liu2016kernelized}
\begin{equation}
 S[q,p] = \argmax_{\boldsymbol{\phi} \in \mathcal{H}^{d}} \{ \mathrm{E}_{q} \left[ trace \left(\mathcal{A}_{p} \boldsymbol{\phi} (\mathbf{m}) \right) \right] ^{2}, \,\,\,\, s.t. \,\,\,\, ||\boldsymbol{\phi}||_{\mathcal{H}^{d}} \leq 1 \}
 \label{eq:kstein_discrepancy}
\end{equation}
where $\mathcal{H}^{d}$ is the RKHS of $d$-dimensional vector functions. The right side of equation (\ref{eq:kstein_discrepancy}) is found to be equal to,
\begin{equation}
\boldsymbol{\phi}^{*} = \boldsymbol{\phi}_{q,p}^{*} ( \mathbf{m} ) / || \boldsymbol{\phi} ^{*}_{q,p} (\mathbf{m}) ||_{\mathcal{H}^{d}} 
\label{eq:phi_star}
\end{equation}
where 
\begin{equation}
 \boldsymbol{\phi} ^{*}_{q,p} (\mathbf{m}) = \mathrm{E}_{\{\mathbf{m'} \sim q\}} [\mathcal{A}_{p} k(\mathbf{m'},\mathbf{m})]
 \label{eq:phi_qp}
\end{equation}
and for which we have $S[q,p] = ||\boldsymbol{\phi} ^{*}_{q,p} (\mathbf{m}) ||_{\mathcal{H}^{d}}$. Thus the optimal $\boldsymbol{\phi}$ in equation (\ref{eq:stein_gradient}) is $\boldsymbol{\phi}^{*}$ and $\nabla_{\epsilon} \mathrm{KL}[q_{T}||p] \, |_{\epsilon=0} = - \sqrt{S[q,p]}$.

Given the above solution, the SVGD works as follows: we start from an initial distribution $q_{0}$, then apply the transform $T^{*}_{0} (\mathbf{m}) = \mathbf{m} + \epsilon \boldsymbol{\phi}^{*}_{ q_{0}, p} (\mathbf{m})$ where we absorb the normalization term in equation (\ref{eq:phi_star}) into $\epsilon$; this updates $q_{0}$ to $q_{[T_{0}]}$ with a decrease in the KL-divergence of $\epsilon*\sqrt{S[q,p]}$. This process is iterated to obtain an approximation of the posterior $p$:
\begin{equation}
 q_{l+1} = q_{l[T^{*}_{l}]}, \,\,\, where \,\,\, T^{*}_{l} (\mathbf{m}) = \mathbf{m} + \epsilon_{l} \boldsymbol{\phi}^{*}_{ q_{l}, p} (\mathbf{m})
 \label{eq:svgd_update} 
\end{equation}
and for sufficiently small $\{\epsilon_{l}\}$ the process eventually converges to the posterior pdf $p$.

\begin{figure}
\includegraphics[width=.5\linewidth]{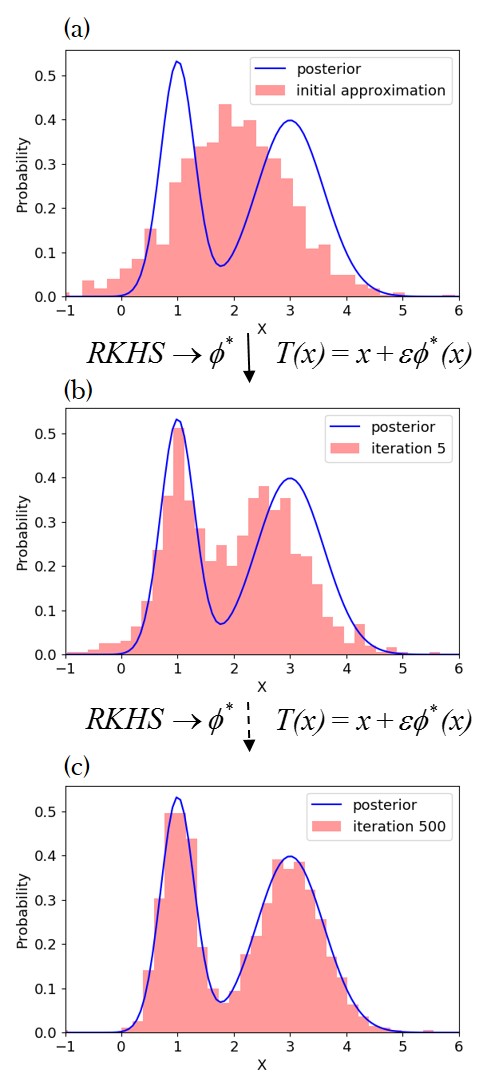}
\caption{An illustration of the SVGD algorithm. The initial pdf is represented by the density of a set of particles (red histogram) in the top plot. The particles are then updated using a smooth transform $T(x)=x+\epsilon\phi^{*}(x)$, where $\phi^{*}$ is found in a reproducing kernel Hilbert space (RKHS). \textbf{(a)} An example of a posterior pdf (blue line) and an initial distribution (red histogram). \textbf{(b)} The approximating probability distribution after 5 iterations. \textbf{(c)} The approximating probability distribution after 500 iterations.}
\label{fig:svgd}
\end{figure}

To calculate the expectation in equation (\ref{eq:phi_qp}) we start from a set of particles (models) generated using $q_{0}$, and at each step the $\boldsymbol{\phi} ^{*}_{q,p} (\mathbf{m})$ can be estimated by computing the mean in equation (\ref{eq:phi_qp}) using those particles. Each particle is then updated using the transform in equation (\ref{eq:svgd_update}), and those particles will form better approximations to the posterior as the iteration proceeds. This suggests the following algorithm which is schematically represented in Figure \ref{fig:svgd}:
\begin{enumerate}[leftmargin=*,labelindent=14pt, align=left]
\item Draw a set of particles $\{ \mathbf{m}_{i}^{0} \}_{i=1}^{n}$ from an initial pdf estimate (e.g., the prior).
\item At iteration $l$, update each particle using:
\begin{equation}
 \mathbf{m}_{i}^{l+1} = \mathbf{m}_{i}^{l} + \epsilon_{l} \boldsymbol{\phi}^{*}_{ q_{l}, p} (\mathbf{m}_{i}^{l})
 \label{eq:particle_update}
\end{equation}
where 
\begin{equation}
 \boldsymbol{\phi}^{*}_{ q_{l}, p} (\mathbf{m}) = \frac{1}{n} \sum_{j=1}^{n} \left[ k(\mathbf{m}_{j}^{l} , \mathbf{m}) \nabla_{\mathbf{m}_{j}^{l}} \mathrm{log} p(\mathbf{m}_{j}^{l}) + \nabla_{\mathbf{m}_{j}^{l}} k(\mathbf{m}_{j}^{l}, \mathbf{m}) \right]
 \label{eq:update_step}
\end{equation}
and $\epsilon_{l}$ is the step size at iteration $l$.
\item Calculate the density of the final set of particles $\{ \mathbf{m}_{i}^{*} \}_{i=1}^{n}$ which approximates the posterior probability density function.  
\end{enumerate}

For kernel $k(\mathbf{m},\mathbf{m'})$ we use the radial basis function $k(\mathbf{m},\mathbf{m'}) = \mathrm{exp}(- \frac{1}{h} ||\mathbf{m}-\mathbf{m'}||^{2}$, where $h$ is taken to be $ \tilde{d}^{2} / \mathrm{log}\, n$ where $\tilde{d}$ is the median of pairwise distances between all particles. This choice of $h$ is based on the intuition that $\sum_{j}k(\mathbf{m}_{i}, \mathbf{m}_{j}) \approx n\mathrm{exp}(-\frac{1}{h} \tilde{d}^{2}) = 1$, so that for particle $\mathbf{m}_{i}$ the two gradient terms in equation (\ref{eq:update_step}) are balanced \cite{liu2016stein}. For the radial basis function kernel the second term in equation (\ref{eq:update_step}) becomes $\sum_{j} \frac{2}{h} (\mathbf{m}-\mathbf{m}_{j}) k(\mathbf{m}_{j}, \mathbf{m})$, which drives the particle $\mathbf{m}$ away from neighbouring particles for which the kernel takes large values. Therefore the second term in equation (\ref{eq:update_step}) acts as a \textit{repulsive force} preventing particles from collapsing to a single mode, while the first term moves particles towards local high probability areas using the kernel-weighted gradient. If in the kernel $h \to 0$, the algorithm falls into independent gradient ascent that maximizes $\mathrm{log} p$ for each particle.

In SVGD, the accuracy of the approximation increases with the number of particles. It has been shown that compared to other particle-based methods, e.g., sequential Monte Carlo methods \cite{smith2013sequential}, SVGD requires fewer samples to achieve the same accuracy which makes it a more efficient method \cite{liu2016stein}. In contrast to sequential Monte Carlo which is a stochastic process, SVGD acts as a deterministic sampling method. If only one particle is used, the second term in equation (\ref{eq:update_step}) becomes zero and the method reduces to a typical gradient ascent towards the model with the maximum a posterior (MAP) pdf value. This suggests that even for a small number of particles the method could still produce a good parameter estimate since MAP estimation can be an effective method in practice.

In seismic tomography velocities are usually constrained to lie within a given velocity range. In order to ensure that velocities always lie within the constrains, we first apply the same transform used in ADVI (equation \ref{eq:transform}) so that the parameters are in an unconstrained space. We can then simply use equation (\ref{eq:particle_update}) to update particles without explicitly considering the constrains on seismic velocities. The final seismic velocities can be obtained by transforming particles back to the constrained space.

\section{Synthetic tests}
\begin{figure}
\includegraphics[width=.9\linewidth]{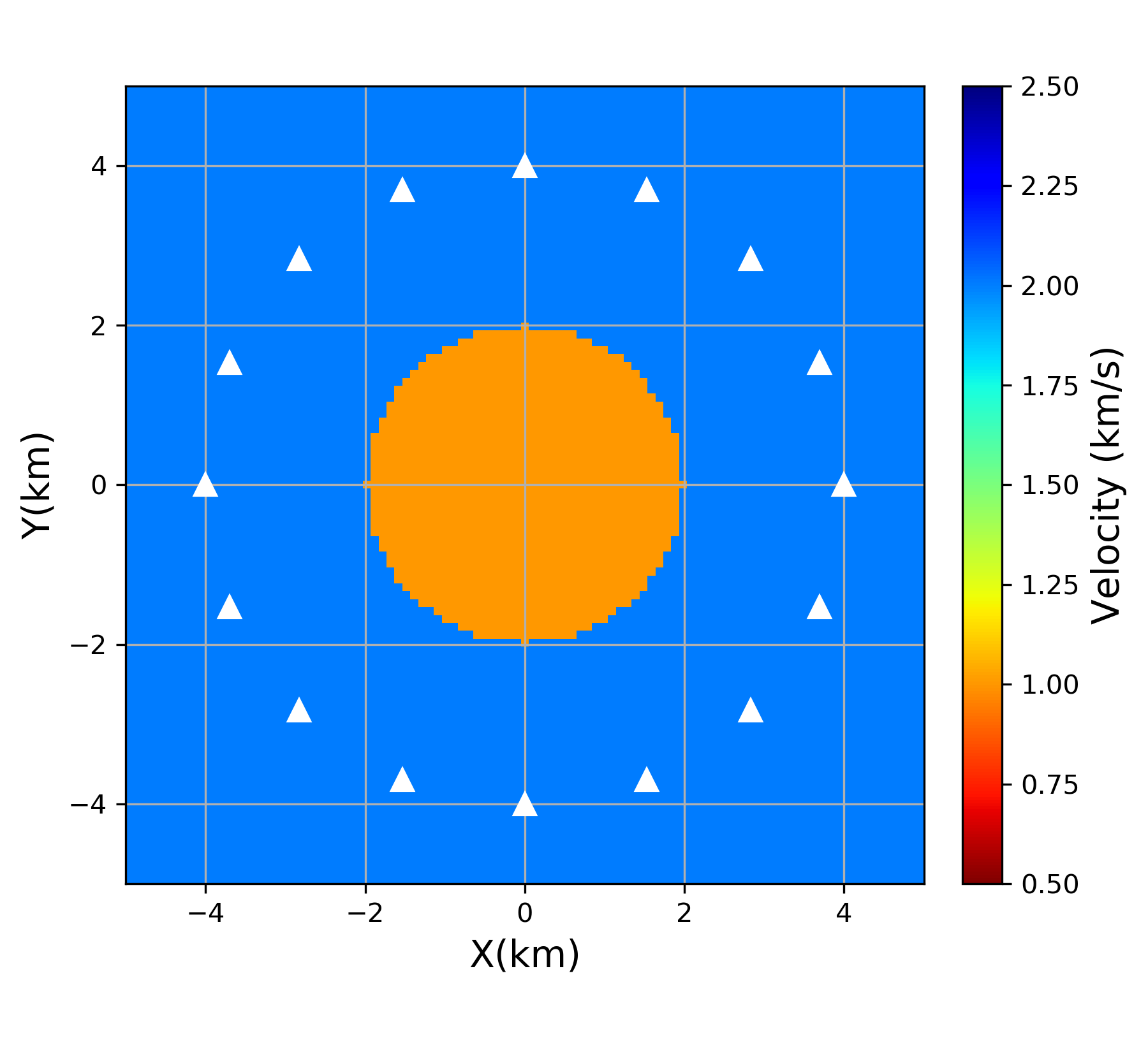}
\caption{The true velocity model and receivers (white triangle) used in the synthetic test. Sources are at the same locations as receivers to simulate a typical ambient noise experiment.}
\label{fig:synthetic}
\end{figure}

We first apply the above methods to a simple 2D synthetic example similar to that in \citet{galetti2015uncertainty}. The true model is a homogeneous background with velocity 2 $km/s$ containing a circular low velocity anomaly with a radius of 2 $km$ with velocity 1 $km/s$. The 16 receivers are evenly distributed around the anomaly approximating a circular acquisition geometry with radius 4 $km$ (Figure \ref{fig:synthetic}). Each receiver is also treated as a source to simulate a typical ambient noise interferometry experiment \cite{campillo2003long, curtis2006seismic, galetti2015uncertainty}. This produces a total of 120 inter-receiver travel time data, each of which is computed using a fast marching method of solving the Eikonal equation over a $100 \times 100$ gridded discretisation in space \cite{rawlinson2004multiple}.

For variational inversions we use a fixed $21 \times 21$ grid of cells to parameterize the velocity model $\mathbf{m}$. The noise level is fixed to be 0.05 $s$ ($<$ 5 percent of travel times) for all inversions. The prior pdf of the velocity in each cell is set to be a Uniform distribution between 0.5 $km/s$ and 3.0 $km/s$ to encompass the true model. Travel times are calculated using the same fast marching method as above over a $100 \times 100$ grid, but using the lower spatial resolution of model properties parameterized in $\mathbf{m}$. The gradients for velocity models are calculated by tracing rays backwards from receiver to (virtual) source using the gradient of the travel time field for each receiver pair \cite{rawlinson2004multiple}. For ADVI, the initial mean of the Gaussian distribution in the transformed space is chosen to be the value which is the transform of the mean value of the prior in the original space, and the initial covariance matrix is simply set to be an identity matrix. We then used 10,000 iterations to update the variational parameters ($\boldsymbol{\mu}$ and $\boldsymbol{\Sigma}$). In order to visualize the results, we generated 5,000 models from the final approximate posterior probability density in the original space and computed their mean and standard deviation. For SVGD, we used 800 particles generated from the prior pdf and transformed to an unconstrained space using equation (\ref{eq:transform}). Each particle is then updated using equation (\ref{eq:svgd_update}) for 500 iterations, then transformed back to seismic velocity. The mean and standard deviation are then calculated using the values of those particles. 

\begin{figure}
\includegraphics[width=1.\linewidth]{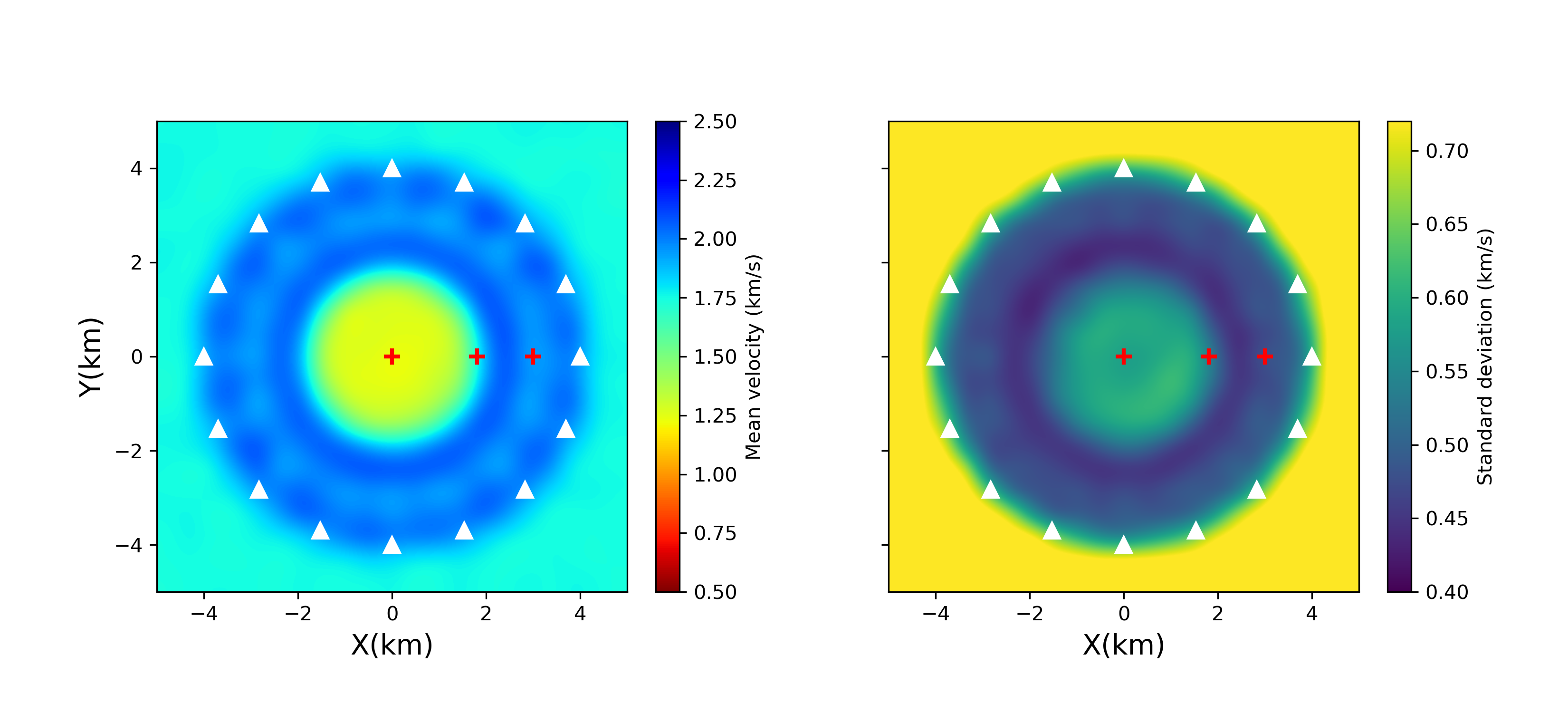}
\caption{The mean (left) and standard deviation (right) found using ADVI. The red pluses show locations which are referred to in the main text.}
\label{fig:synthetic_advi}
\end{figure}

To demonstrate the variational methods we compare the results with the fixed-dimensional Metropolis-Hastings McMC (MH-McMC) method \cite{metropolis1949monte, hastings1970monte, mosegaard1995monte, malinverno2000monte} and the rj-McMC method \cite{green1995reversible, bodin2009seismic, galetti2015uncertainty, zhang20183}. For MH-McMC inversion we used the same parameterization as for the variational methods (a $21 \times 21$ grid). A Gaussian perturbation is used as the proposal distribution used to generate potential McMC samples, for which the step length is chosen by trial and error to give an acceptance ratio between 20 and 50 percent. We used a total of 6 chains, each of which used 2,000,000 iterations with a burn-in period of 1,000,000 iterations. To reduce the correlation between samples we only retain every 50$^{th}$ sample in each chain after the burn-in period. The mean and standard deviation are then calculated using those samples. For rj-McMC inversion we use Voronoi cells to parameterize the model \cite{bodin2009seismic}, for which the prior pdf of the number of cells is set to be a Uniform distribution between 4 and 100. The proposal distribution for fixed-dimensional steps (changing the velocity of a cell or moving a cell) is chosen in a similar way as in MH-McMC. For trans-dimensional steps (adding or deleting a cell) the proposal distribution is chosen as the prior pdf \cite{zhang20183}. We used a total of 6 chains, each of which contained 500,000 iterations with a burn-in period of 300,000. Similarly to the fixed-dimensional inversion the chain was thinned by a factor of 50 post burn-in.  

\begin{figure}
\includegraphics[width=1.\linewidth]{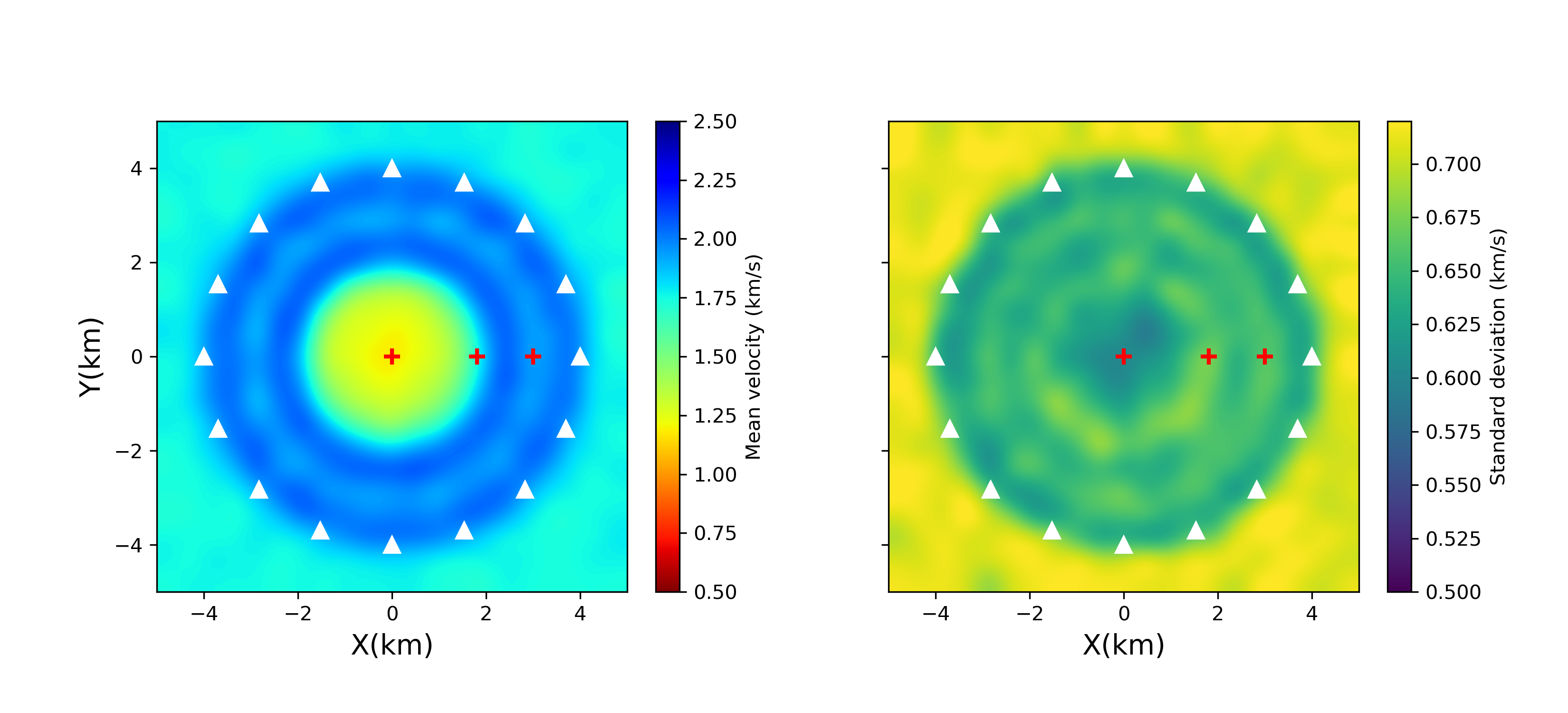}
\caption{The mean (left) and standard deviation (right) found using SVGD. The red pluses show locations which are referred to in the main text.}
\label{fig:synthetic_svgd}
\end{figure}

\subsection{Results} 
Figure \ref{fig:synthetic_advi} shows the mean and standard deviation calculated using ADVI. The mean model successfully recovers the low velocity anomaly within the receiver array except that the velocity value is slightly higher ($\sim 1.2 \, km/s$) than the true value ($1.0 \, km/s$). Between the location of the central anomaly and that of the receiver array there is a slightly lower velocity loop. The standard deviation map shows standard deviations similar to that of the prior (0.72 km/s) outside of the array, and clearly higher uncertainties at the location of the central anomaly. The standard deviations around the central anomaly are slightly higher than those at the center. Figure \ref{fig:synthetic_svgd} shows the results from SVGD. Similarly, the velocity of the low velocity anomaly ($\sim 1.2 \, km/s$) is slightly higher than the true value and a slightly lower velocity loop is also observed between the central anomaly and the receiver array. There is a clear higher uncertainty loop around the central anomaly; this has been observed previously and represent uncertainty due to the trade-off between the velocity of the anomaly and its shape \cite{galetti2015uncertainty, zhang20183}. There is also another higher uncertainty loop associated with the lower velocity loop between the central anomaly and the reciever array. In contrast to this result, the loop cannot be observed in the results of ADVI. 

\begin{figure}
\includegraphics[width=1.\linewidth]{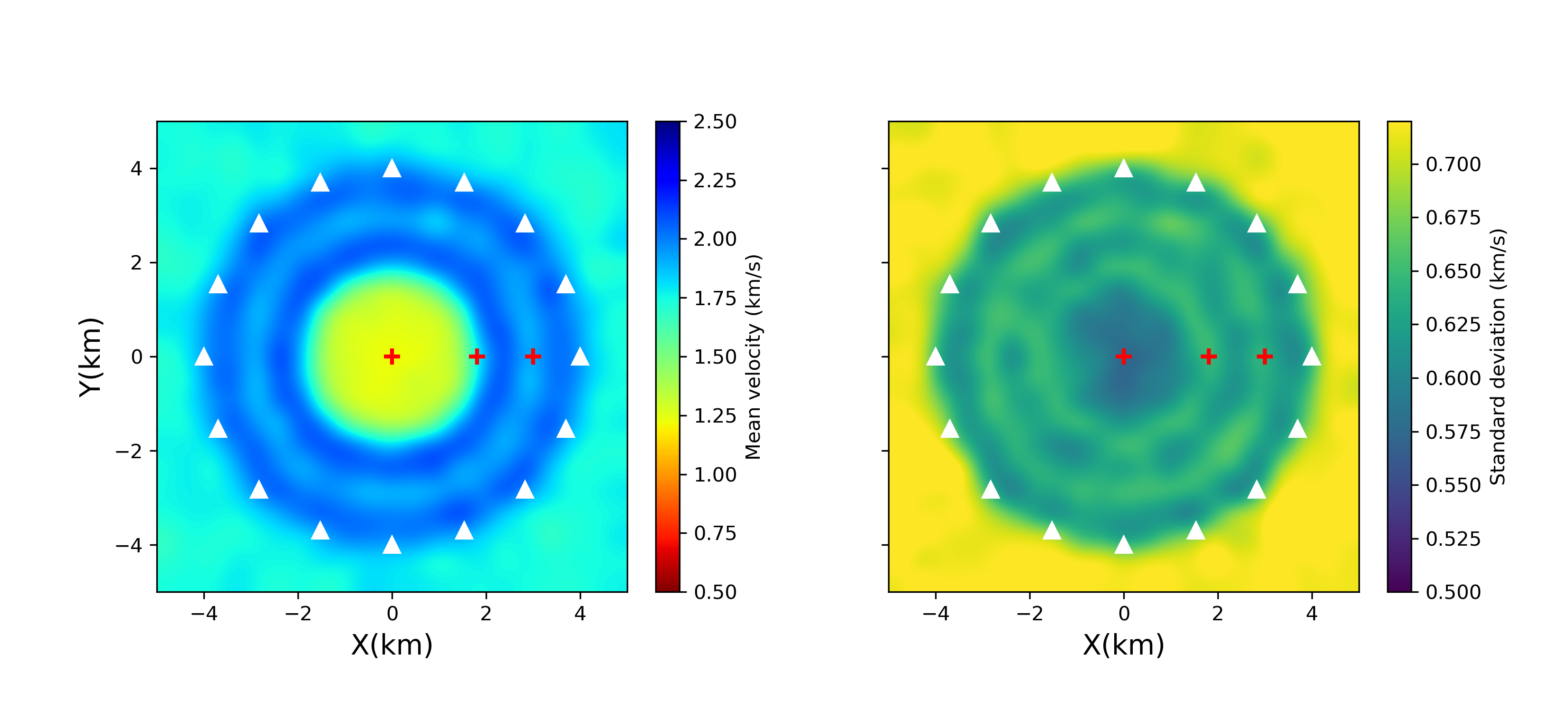}
\caption{The mean (left) and standard deviation (right) found using MH-McMC. The red pluses show the point location which are referred to in the text.}
\label{fig:synthetic_mcmc}
\end{figure}

\begin{figure}
\includegraphics[width=1.\linewidth]{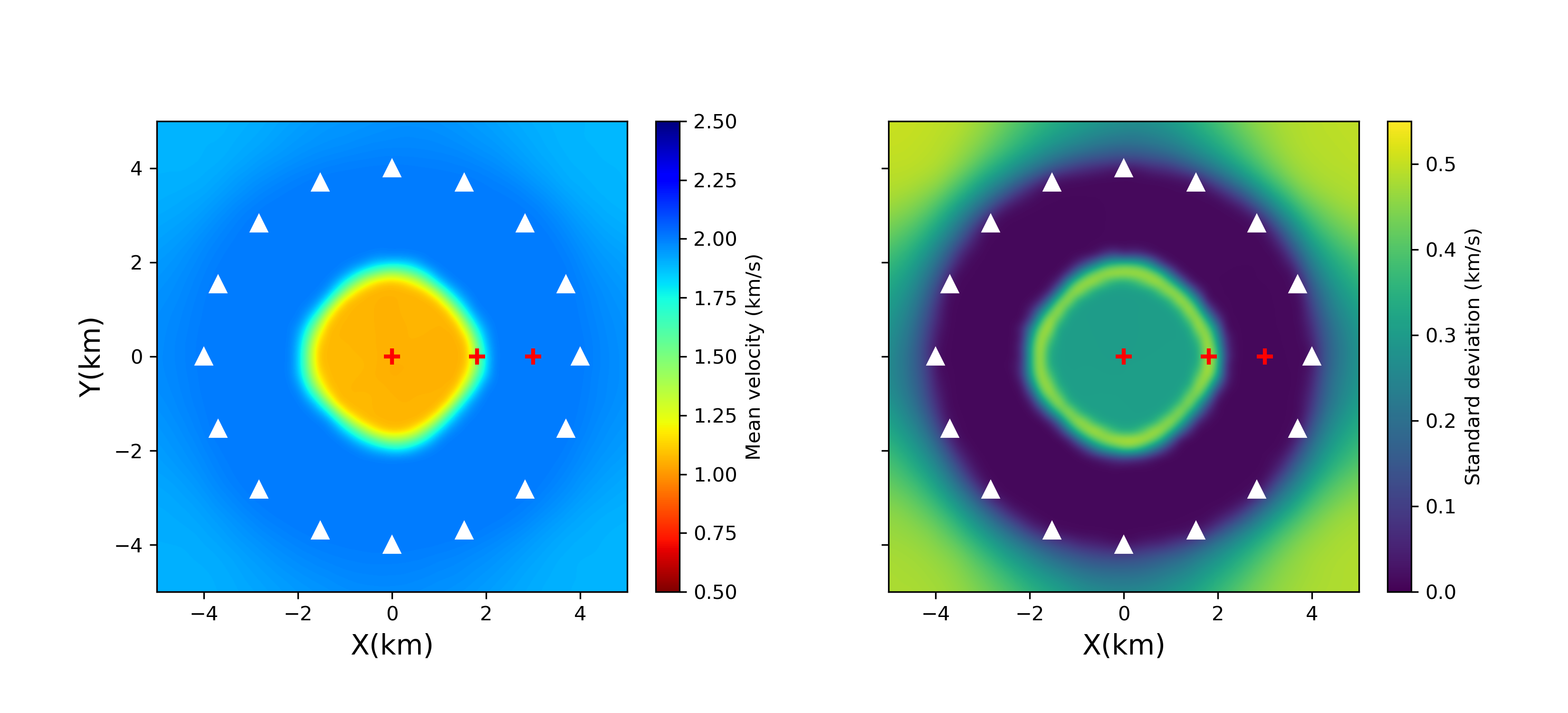}
\caption{The mean (left) and standard deviation (right) found using trans-dimensional rj-McMC. The red pluses show the point location which are referred to in the text.}
\label{fig:synthetic_rjmcmc}
\end{figure}

To validate and better understand these results, Figure \ref{fig:synthetic_mcmc} shows the results from MH-McMC. The mean velocity model is very similar to the results from ADVI and SVGD. For example, the velocity value of the low velocity anomaly is higher than the true value, which suggests that the mean value of the posterior under the specified parameterization is genuinely biased towards higher values than the true value. A lower velocity loop is also observed between the circular anomaly and the receiver array. The standard deviation map shows similar results to those from SVGD: there is a higher uncertainty loop around the central anomaly and another one associated with the lower velocity loop between the circular anomaly and the receiver array. The latter loop suggests that this area is not well constrained by the data, and therefore the mean velocity tends towards the mean value of the prior which is lower than the true value. We do not observe the clear higher uncertainty loops in the result of ADVI which may be due to the Gaussian approximation which is used to fit a non-Gaussian posterior. In Figure \ref{fig:synthetic_rjmcmc} we show the results from rj-McMC. Compared to the results from the fixed-parameterization inversions, the mean velocity is a more accurate estimate of the true model and uncertainty across the model is also lower. For example, the middle low velocity anomaly has almost the same value as the true model and has standard deviation of only $\sim 0.3 \, km/s$ compared to values significantly greater than 0.3 $km/s$ for all other methods. Between the middle anomaly and the receivers, the model is determined better than in the fixed-paramterization inversions (with a standard deviation smaller than $0.1 \, km/s$). This is because in rj-McMC the model parameterization adapts to the data which usually results in a lower-dimensional parameter space due to the natural parsimony of the method. For example, the average dimensionality of the parameter space in the rj-McMC inversion is around 10; for comparison the fixed-parameterization inversions all have dimensionality fixed to be 441. The standard deviation map from the rj-McMC also shows a clear higher uncertainty loop within the array around the low velocity anomaly, and high uncertainties outside of the array where there is no data coverage. 

The results in Figure \ref{fig:synthetic_rjmcmc} do not show the double-loop uncertainty structure that is observed in the SVGD and MH-McMC results. The rj-McMC method contains an implicit natural parsimony -- the method tends to use fewer rather than more cells whenever possible. While this may be useful in order to reduce the dimensionality of parameter space, it is also possible that it causes some detailed features of the velocity or uncertainty structure to be omitted, much like a smoothing regularization condition in other tomographic methods. Since the double-loop structure appears to be a robust feature of the image uncertainty, we assume that the parsimony has indeed regularised some of the image structure out of the rj-McMC results.

\begin{figure}
\includegraphics[width=1.\linewidth]{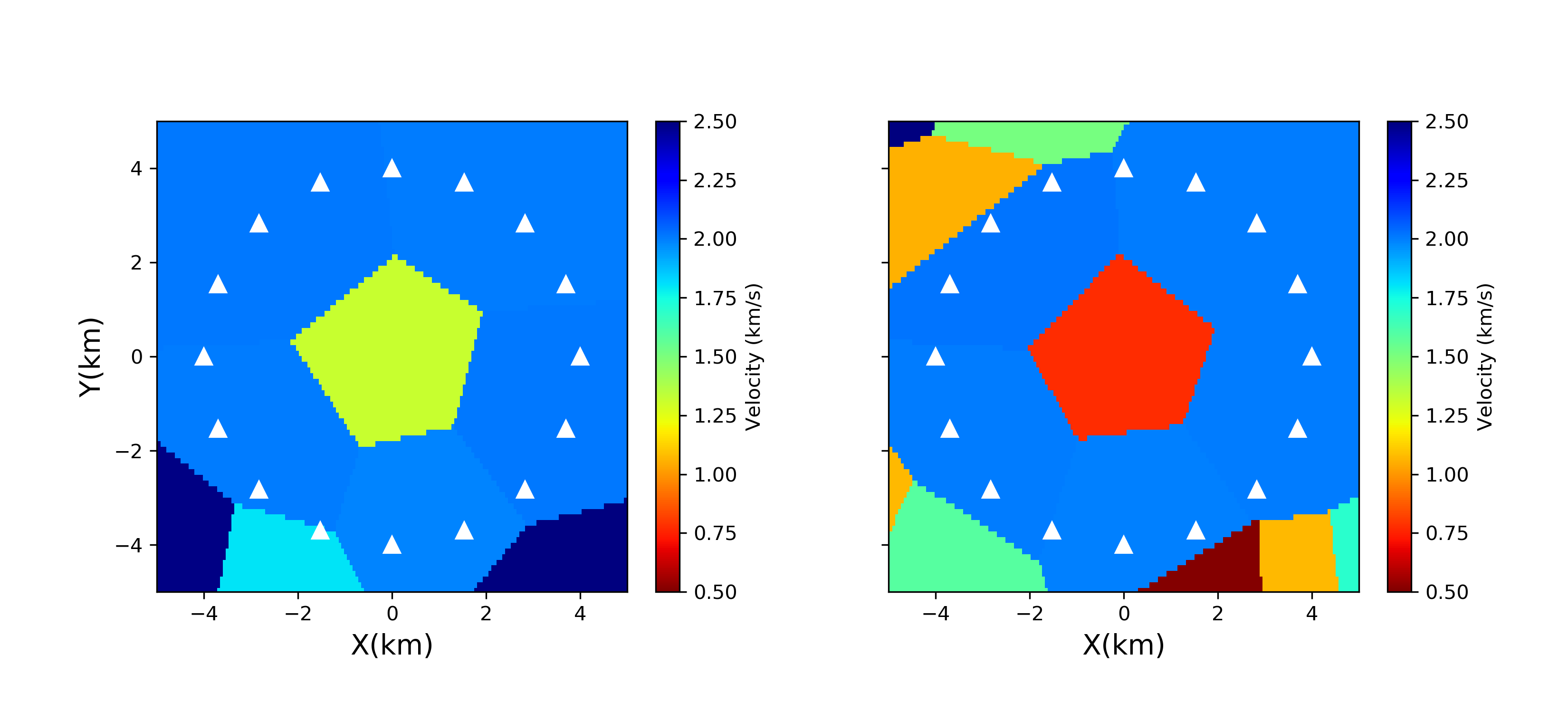}
\caption{Two examples of the models sampled using rj-McMC algorithm.}
\label{fig:synthetic_rjmcmc_models}
\end{figure}

Note that the result from rj-McMC is fundamentally different from results obtained using the fixed-parameterization inversions (ADVI, SVGD and MH-McMC) because of its entirely different parameterization. While the other inversion results can themselves be regarded as pixelated images, rj-McMC produces a set of models that are not images (Figure \ref{fig:synthetic_rjmcmc_models}). The results shown in Figure \ref{fig:synthetic_rjmcmc} are pointwise means and standard deviations of velocities implied by a set of models similar to nature to those in Figure \ref{fig:synthetic_rjmcmc_models} so that images are presented in a very different space to that of the model parameters. 
 
\begin{figure}
\includegraphics[width=1.\linewidth]{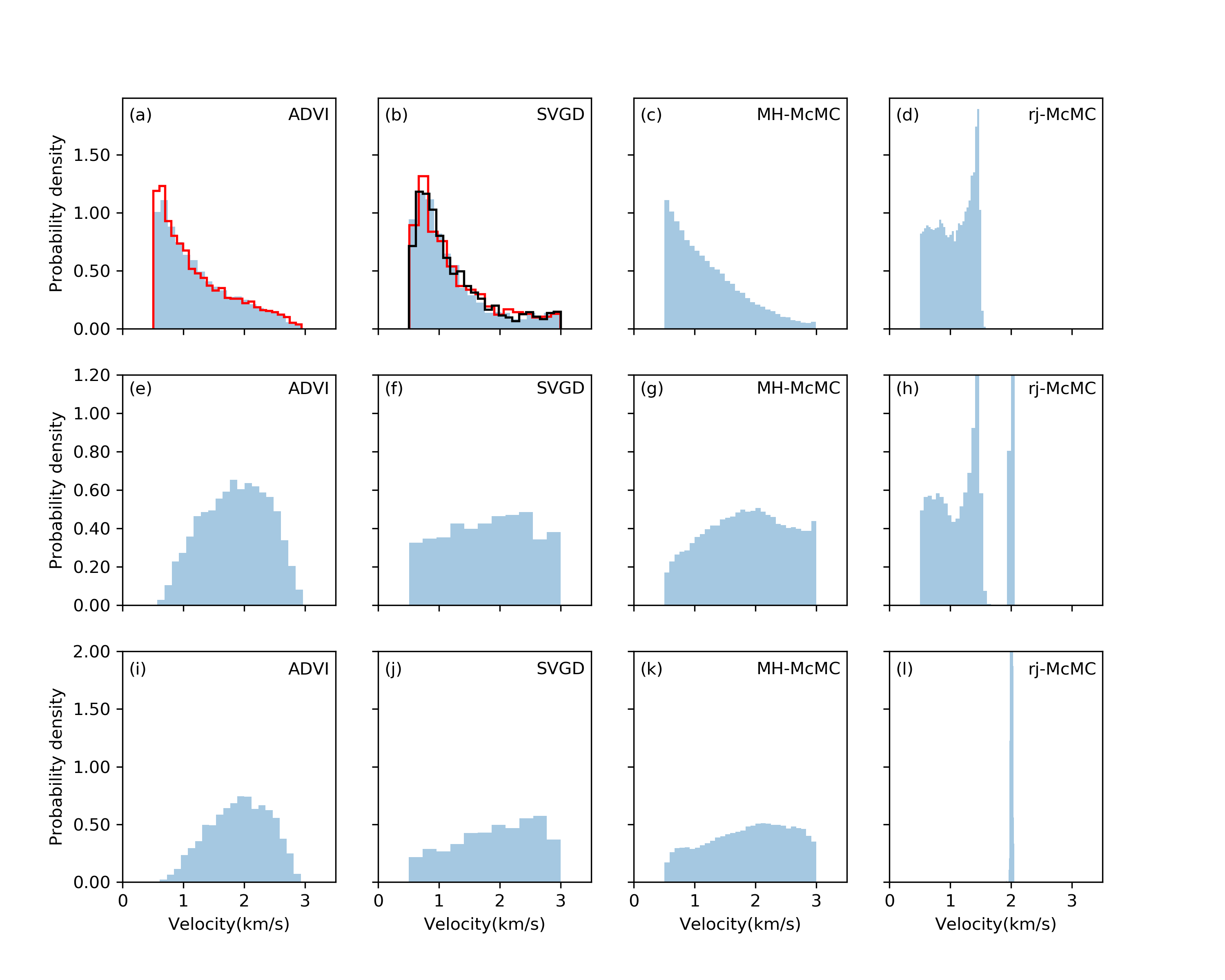}
\caption{The marginal posterior pdfs of velocity at three points (pluses in Figure 3,4,5,6) derived using different methods. \textbf{(a)}, \textbf{(b)}, \textbf{(c)} and \textbf{(d)} show the marginal posterior distributions of velocity at the point (0,0) from ADVI, SVGD, MH-McMC and rj-McMC respectively. \textbf{(e)}, \textbf{(f)}, \textbf{(g)} and \textbf{(h)} show the marginal distributions at the point (1.8,0) from the four methods respectively, and \textbf{(i)}, \textbf{(j)}, \textbf{(k)} and \textbf{(l)} show the marginal distributions at the point (3,0) from the four methods respectively. The red lines in \textbf{(a)} and \textbf{(b)} are marginal distributions obtained by doubling the number of iterations and the black line in \textbf{(b)} shows the marginal distribution obtained using 1,600 particles.} 
\label{fig:points_density}
\end{figure}

To further analyse the results, in Figure \ref{fig:points_density} we show marginal probability distributions from the different inversion methods at three points (plus signs in Figure \ref{fig:synthetic_advi}, \ref{fig:synthetic_svgd}, \ref{fig:synthetic_mcmc}, and \ref{fig:synthetic_rjmcmc}): point (0, 0) at the middle of the model, point (1.8, 0) at the boundary of the low velocity anomaly which has higher uncertainties, and point (3, 0) which also has higher uncertainties in the results from SVGD and MH-McMC. Due to symmetries of the model, marginal distributions at these three points are sufficient to reflect much of the entire set of single-parameter marginal probability distributions. At point (0, 0), the three fixed-parameterization methods produce similar marginal probability distributions. However, the marginal distribution from rj-McMC is narrower and concentrates around the true solution ($1.0 \, km/s$). This is likely due to the fact that in rj-McMC we have a much smaller parameter space than in the fixed-parameterization inversions. To assess the convergence we show the marginal distributions obtained by doubling the number of iterations in ADVI and SVGD with an red line in Figure \ref{fig:points_density}a and b. The results show that increasing iterations only slightly improves the marginal distributions, suggesting that they have nearly converged. The black line in Figure \ref{fig:points_density}b shows the marginal distribution obtained using more particles (1,600) with the same number of iterations (500). The result is almost the same as the result obtained using the original set of particles which suggests that 800 particles are sufficient in this case. At point (1.8, 0), the marginal distributions from the three fixed-parameterization inversions become broader which explains the higher uncertainty loops observed in the standard deviation maps. The distribution from ADVI is more centrally focussed than the other two, which is again suggestive of the limitations of that method caused by the Gaussian approximation. The distributions from SVGD and MH-McMC are more similar to each other and are close to the prior -- a Uniform distribution -- which suggests that the area is not well constrained by the data. By contrast, the result from rj-McMC shows a clearly multimodal distribution with one mode centred around the velocity of the anomaly (1 $km/s$) and the other around the background velocity (2 $km/s$) as discussed in \citet{galetti2015uncertainty}. This multimodal distribution reflects the fact that it is not clear whether this point is inside or outside of the anomaly which produces the higher uncertainty loop in the standard deviation map. This suggests that there are different causes of the higher uncertainty loops in the different models. In the fixed-parameterization inversions (ADVI, SVGD and MH-McMC) the higher uncertainty loops are mainly caused by the low resolution of the data at the boundary of the low velocity anomaly which produces broader marginal distributions. In the rj-McMC inversion, the higher uncertainty loops are mainly caused by multimodality in the posterior pdf. At point (3.0, 0) similarly to the point (0, 0), the marginal distributions from the three fixed-parameterization inversions have similar shape and are much broader than the result from rj-McMC. Compared to the results from SVGD and MH-McMC, the result from ADVI again shows a more centrally-focussed distribution reminiscent of the Gaussian limitation implicit in ADVI. In the result of rj-McMC the marginal distribution concentrates to a very narrow distribution around the true value. Overall the marginal distributions from the fixed-parameterization inversions are broader than the result from rj-McMC due to their far larger parameter space. Note that although the marginal distributions from SVGD and MH-McMC have slightly different shape which causes differences in the magnitudes of their standard deviation maps, the maps are essentially similar from these quite different methods which suggests that the results are (approximately) correct. 
     
\subsection{Computational cost}
Table 1 summarises the computational cost of the different methods. ADVI involves 10,000 forward simulations which takes 0.45 CPU hours. However, note that in ADVI we used the full-rank covariance matrix which becomes huge in high dimensional parameter spaces which could makes the method inefficient. SVGD involves 400,000 forward simulations which takes 8.53 CPU hours. This appears to make it less efficient than ADVI, however SVGD can produce a more accurate approximation to the posterior pdf than ADVI which is limited by the Gaussian approximation. Note that SVGD can easily be parallelized by computing the gradients in equation (\ref{eq:update_step}) in parallel, making the method more time-efficient. For example, the above example takes 0.97 hours when parallelized using 10 cores. In comparison, MH-McMC requires 2,000,000 simulations for one chain which takes about 80.05 CPU hours, so for all 6 chains it requires 480.3 CPU hours in total. The rj-McMC run involved 500,000 simulations for one chain which takes about 17.1 CPU hours, so 102.6 CPU hours in total for 6 chains. The Monte Carlo methods use evaluations of the likelihood and prior distribution at each sample whereas both variational methods also deploy the information in the various gradients in equations \ref{eq:gradient_mu}, \ref{eq:gradient_sigma} and \ref{eq:update_step}. The number of simulations is therefore not a good metric to compare the four methods, since the gradients in this case are calculated by ray tracing which require more calculations per simulation in Table 1 compared to MC. CPU hours is a fairer metric for comparison, but of course this depends on the mechanism by which gradients are obtained: in other forward or inverse problems it is even possible that the variational methods take longer than Monte Carlo if estimating gradients requires extensive computation.

In the comparison in Table 1, rj-McMC is more efficient than MH-McMC due to the fact that rj-McMC explores a much smaller parameter space than the fixed parameterization in MH-McMC. However, note that this might not always be true since trans-dimensional steps in rj-McMC usually have a very low probability of being accepted \cite{bodin2009seismic, zhang20183} and the method is generally significantly more difficult to tune \cite{green2009reversible}. Overall, obtaining solutions from variational methods (ADVI, SVGD) is more efficient than Monte Carlo methods since they turn the Bayesian inference problem into an optimization problem. This also makes variational inference methods applicable to larger-datasets, and offers the advantage that very large datasets can be divided into random minibatches and inverted using stochastic optimization \cite{robbins1951stochastic, kubrusly1973stochastic} together with distributed computation. Monte Carlo methods are very computationally expensive for large datasets. Of course, the above comparison depends on the methods used to assess convergence for each method, which introduces some subjectivity in the comparison so that the absolute time required by each method may not be entirely accurate. Nevertheless, from all tests that we have conducted it is clear that variational methods produce solutions far more efficiently than Monte Carlo methods.

 \begin{table}
 \caption{The comparison of computational cost for all 4 methods}
 \centering
 \begin{tabular}{l c c}
 \hline
  Method  & Number of simulations & CPU hours  \\
 \hline
   ADVI  & 10,000 & 0.45   \\
   SVGD & {400,000} & 8.53 \\
   MH-McMC & 12,000,000 & 480.3 \\
   rj-McMC & 3,000,000 & 102.6 \\
 \hline
 \end{tabular}
 \end{table}

\section{Application to Grane field}
\begin{figure}
\includegraphics[width=.8\linewidth]{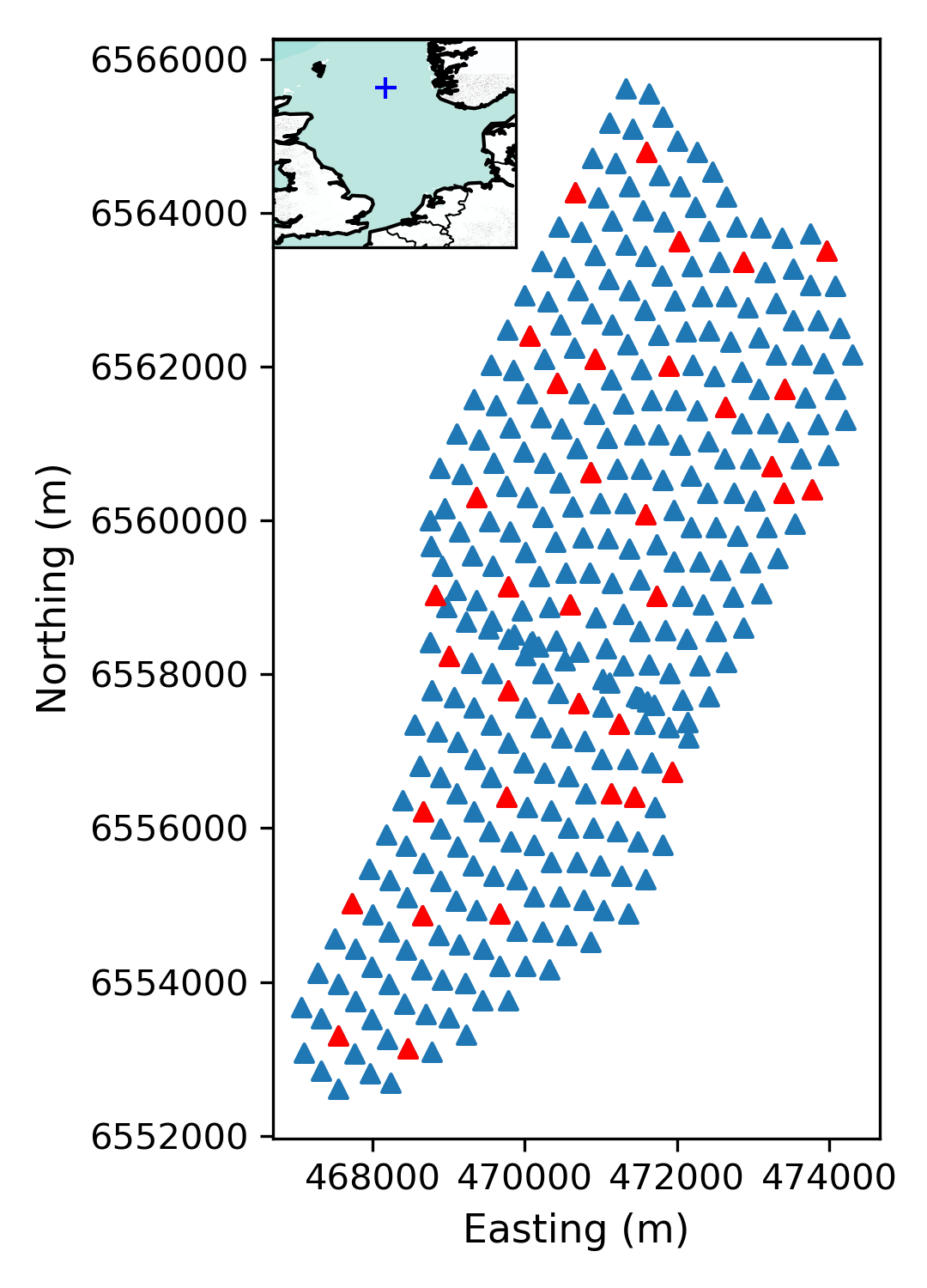}
\caption{The distribution of receiver (blue and red triangles) across the Grane field used in this study. Red triangles show the receivers that were used as virtual sources. The blue plus in the inset map shows the location of Grane field.}
\label{fig:Grane_receivers}
\end{figure}

The Grane field is situated in the North sea, and contains a permanent monitoring system composed of 3458 four-component sensors measuring 3 orthogonal components of particle velocity and water pressure variations due to passing seismic waves. This allows us to use ambient seismic noise tomography to study the subsurface of the field. To reduce the computational cost, in this study we down-sampled the number of receivers by a factor of 10 which results in 346 receivers, and we only used 35 receivers as virtual sources (Figure \ref{fig:Grane_receivers}). Cross-correlations are computed between vertical component recordings at pairs consisting of a virtual source and a receiver using half-hour time segments, and the set of correlations for each pair were stacked over 6.5 hours. This process produces approximate virtual-source seismograms of Rayleigh-type Scholte waves \cite{campillo2003long, shapiro2005high, curtis2006seismic}. Phase velocity dispersion curves for each (virtual) source-receiver pair are then automatically picked using an image transformation technique: for all processing details see \citet{zhang2019fully} which presents a complete ambient noise analysis of the field and presents tomographic phase velocity maps at various frequencies as well as estimated shear-velocity structure of the near seabed subsurface. Here we use the recording phase velocity data at 0.9s period.  

We apply the variational inference methods ADVI and SVGD, and rj-McMC to the data to obtain phase velocity maps at 0.9 s and compare the results. For variational methods, the field is parametrized using a regular $26 \times 71$ grid with a spacing of 0.2 km at both x and y directions giving a velocity model dimensionality of 1846. Due to its computational cost in high dimensional spaces we do not apply MH-McMC. The data noise level is set to be 0.05 s, which is an average value estimated by the hierarchical Bayesian Monte Carlo inversion of \citet{zhang2019fully}. The prior pdf of phase velocity in each model cell is set to be a Uniform distribution between $0.35\, km/s$ and $0.55\, km/s$, which is selected to be wider than the minimum ($0.4 \,km/s$) and maximum ($0.5 \,km/s$) phase velocity picked from cross-correlations. We then applied 10,000 iterations for ADVI and for SVGD we used 1000 particles and 500 iterations. Similarly to the synthetic test above for rj-McMC we use Voronoi cells to parameterize the model. The prior pdf of the number of cells is set to be a discrete Uniform distribution between 30 and 200, and the data noise level is estimated hierarchically during the inversion \cite{zhang20183}. Proposal distributions are the same as in the synthetic test above. We used a total of 16 chains, each of which contains 800,000 iterations including a burn-in period of 400,000. To reduce the correlation between samples we only retain every 50\textsuperscript{th} sample post burn-in for our final ensemble.
  
\begin{figure}
\includegraphics[width=1.\linewidth]{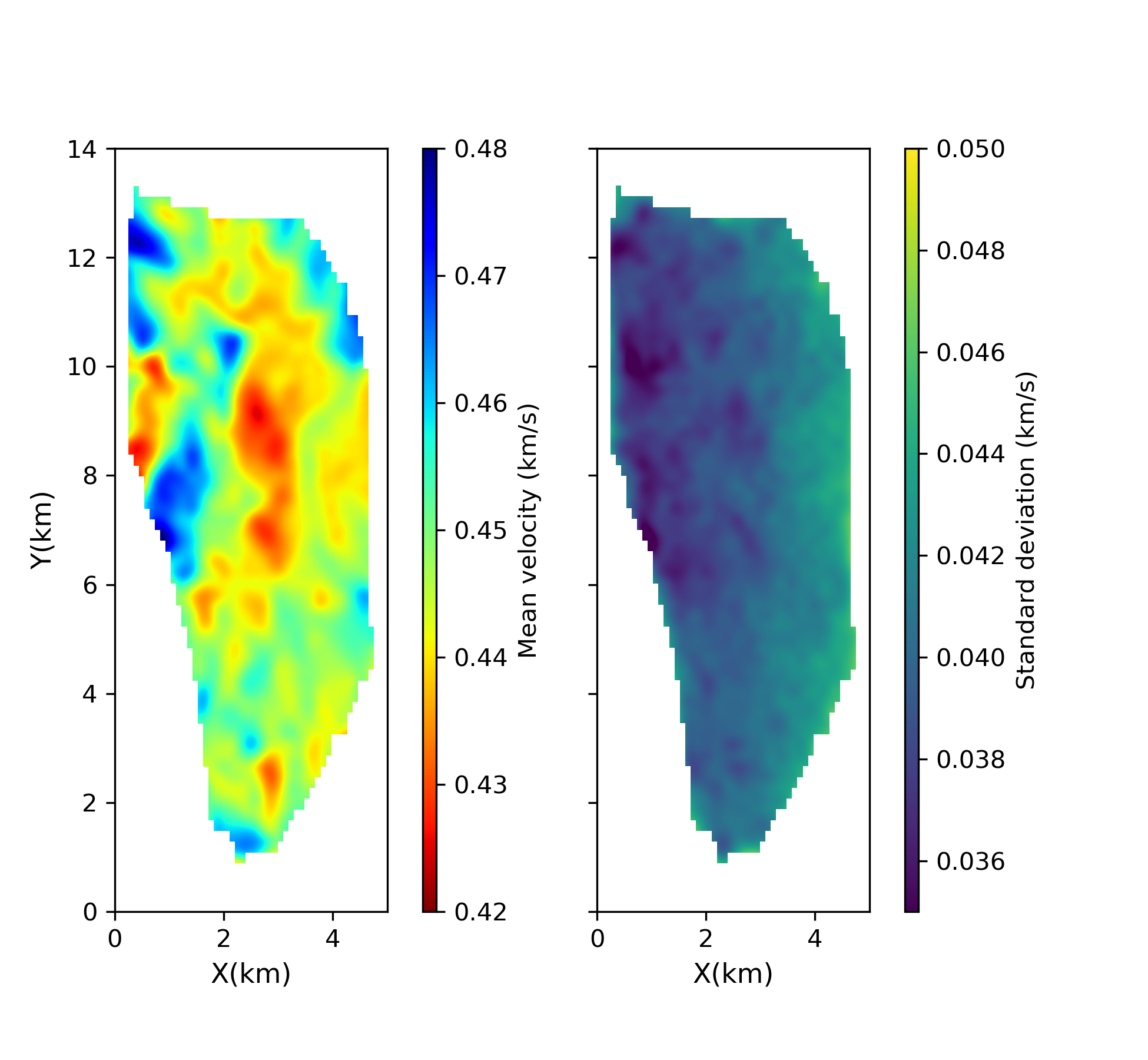}
\caption{The mean (left) and standard deviation map (right) from ADVI.}
\label{fig:Grane_advi}
\end{figure}

Figure \ref{fig:Grane_advi} shows the mean and standard deviation maps from ADVI. The mean phase velocity map shows a clear low velocity anomaly around the centre of the field from Y=6 km to Y=10 km and another at the western edge between Y=8 km and Y=10 km. These were also observed by \cite{zhang2019fully} using Eikonal tomography, who showed that they are correlated with areas of higher density of pockmarks on the seabed, suggesting that they are caused by near surface fluid flow effects. At the western edge between Y=6 km and Y=8 km and at the northwestern edge there are high velocity anomalies which were also observed in the results of \citet{zhang2019fully}. In the north between Y=11 km and Y=12 km and along the eastern edge between Y=7 km and Y=10 km the model shows some low velocity anomalies. Moreover, there are some small anomalies distributed across the field. For example, to the south of the central low velocity anomaly around Y=6 km there are several other low velocity anomalies. Similarly there is a small low velocity anomaly and a small high velocity anomaly in the south of the field around Y=2.5 km, and a small high velocity anomaly in the north around Y=10.5 km.
   
Overall the standard deviation map shows that uncertainty in the west is lower than in the east. At the western edge there are some low uncertainty areas which are associated with velocity anomalies. For example, the low uncertainty area between Y=6 km and Y=8 km is associated with the high velocity anomaly at the same location. Similarly the high velocity anomaly at the northwestern edge around Y=12 km shows a lower uncertainty, and the middle low velocity anomaly also shows slightly lower uncertainties. This might suggest that these velocity structures are well-constrained by the data. However, in the synthetic tests we noticed that the ADVI can produce biased standard deviation maps due to the Gaussian approximation, so these uncertainty properties may not be robust.

\begin{figure}
\includegraphics[width=1.\linewidth]{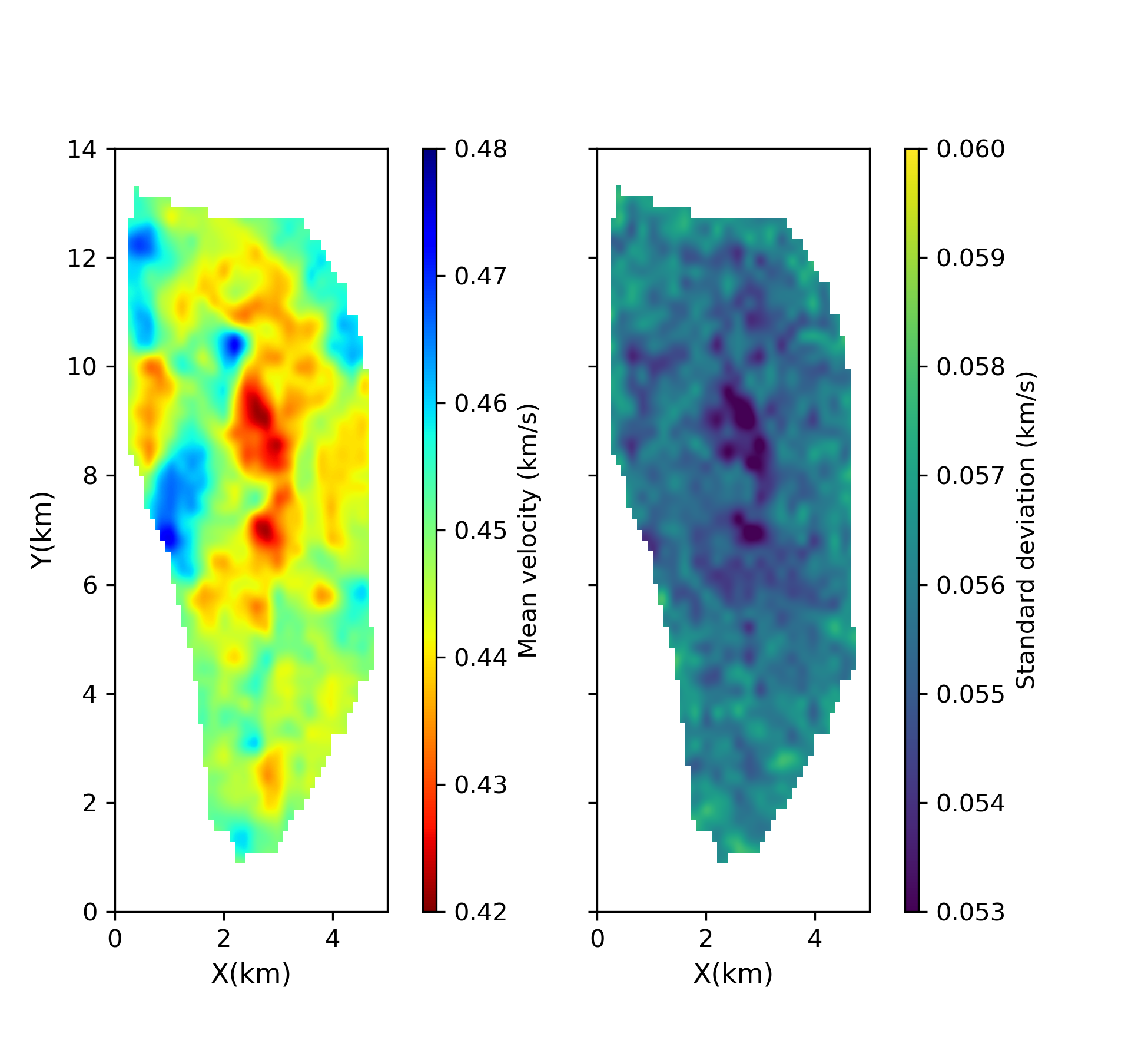}
\caption{The mean (left) and standard deviation map (right) from SVGD.}
\label{fig:Grane_svgd}
\end{figure}

\begin{figure}
\includegraphics[width=1.\linewidth]{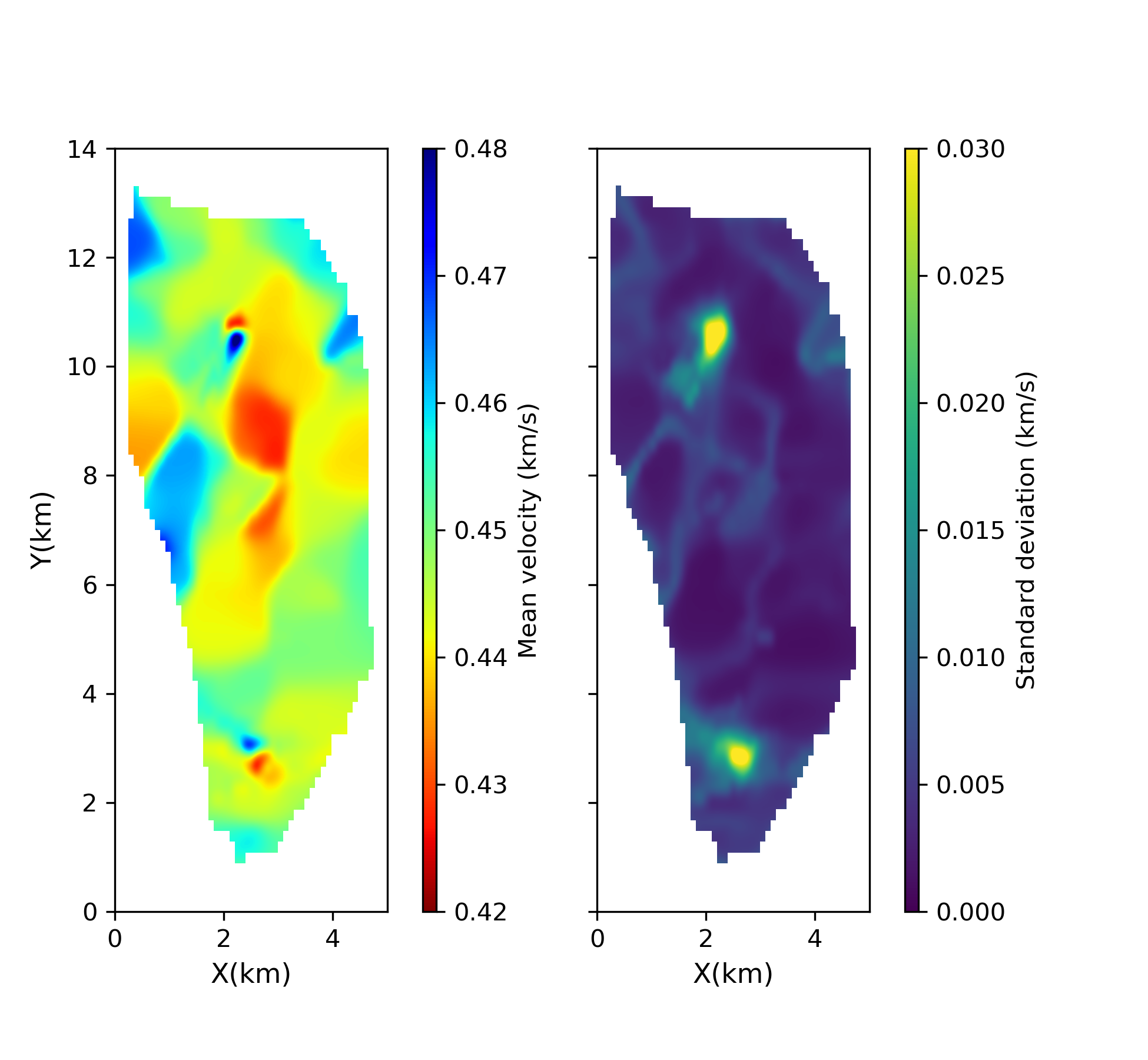}
\caption{The mean (left) and standard deviation map (right) from rj-McMC.}
\label{fig:Grane_rjmcmc}
\end{figure}

We show the mean and standard deviation maps obtained using SVGD in Figure \ref{fig:Grane_svgd}. The mean velocity map shows very similar structures to the result from ADVI, except that the velocity magnitudes are slightly different. For example, we observe the central low velocity anomaly and one at the western edge which appeared in the mean velocity map from ADVI and are related to the density distribution of pockmarks. Similarly there are high velocity anomalies at the western edge and a low velocity anomaly at the eastern edge. Even for more detailed structure, e.g., the low velocity anomalies at the north (Y > 10 km), the low velocity anomalies around Y=6 km and the small velocity anomalies around Y=2.5 km, the two results show highly consistent properties between the two methods. This suggests that we have obtained accurate mean phase velocity maps given the fixed, gridded model parameterization and the observed data.

Despite the similarity in the mean results, the standard deviation map from SVGD is quite different from the results from ADVI, which is consistent with similar variations that we observed in the synthetic tests. For example, there is no clear magnitude difference between the west and the east as appeared in the result from ADVI. There is a clear low uncertainty area associated with the central low velocity anomaly, which is slightly lower in magnitude than the result from ADVI. Similarly there is a slightly lower uncertainty area at the western edge associated with the low velocity anomaly at the same location. The south-central low velocity anomaly around Y=6 km also exhibits relatively lower uncertainties, which suggests that those small low velocity anomalies in this area may reflect true properties of the subsurface. Similarly there are some low uncertainty structures at the north around Y= 11 km which are associated with low velocity anomalies. Note that due to the Gaussian approximation in ADVI, the standard deviation results from SVGD show different magnitudes as we saw in the synthetic tests.

Figure \ref{fig:Grane_rjmcmc} shows the mean and standard deviation maps obtained from rj-McMC. The mean velocity map shows broadly similar structures to the results from ADVI and SVGD. For example, we also observed the middle low velocity anomaly, the low velocity anomalies at the western and eastern edges and the high velocity anomalies at the western edge. However, compared to the previous results these structures are smoother which is probably caused by the natural parsimony that is implicit within the rj-McMC inversion method \cite{green1995reversible, bodin2009seismic} similarly to the synthetic tests above. The small velocity anomalies in the previous results disappear in the result from rj-McMC; this may also be caused by the natural parsimony of rj-McMC, or by overfitting of data in the variational methods due to the fixed parameterization. However, the small high and low velocity anomalies around Y=2.5 km and around Y=10.5 km still exist, which suggests that these detailed velocity structures may represent real properties of the subsurface (or are caused by a consistent bias in the data).

Similarly to the synthetic tests, the standard deviation map from rj-McMC shows significantly smaller uncertainties ($< 0.01$ km/s) than the results from ADVI ($\sim 0.04$ km/s) and SVGD ($\sim 0.055$ km/s), which is probably caused by a lower dimensionality of parameter space used in rj-McMC (around 60 Voronoi cells were used) than in variational methods ($1846$), resulting in fewer trade-offs between parameters. However, there are higher uncertainties at the location of the small velocity anomalies at Y=2.5 km and at Y=10.5 km, which is probably due to the fact that not all chains found these small structures. 

In the inversion, ADVI involved 10,000 forward simulations which took 5.1 CPU hours and SVGD involved 500,000 forward simulations which required 141.8 CPU hours. By contrast the rj-McMC involved 12,800,000 forward simulations to obtain an acceptable result which required 1,866.1 CPU hours. In real time, SVGD was in fact parallelised using 12 cores which took 12.1 hours to run, while rj-McMC was parallelised using 16 cores which therefore took about 5 days. We conclude that, although the variational methods produce higher uncertainty estimates, they can produce similar parameter estimates (mean velocity) at hugely reduced computational cost, and indeed our synthetic tests suggest that the variational SVGD image uncertainty results may in fact be more correct.
 
\section{Discussion}
We have shown that variational methods (ADVI and SVGD) can be applied to seismic tomography problems and provide efficient alternatives to McMC. ADVI produces biased posterior pdfs because of its implicit Gaussian approximation. However, it still generates an accurate estimate of the mean model. Given that it is very efficient (only requiring 10,000 forward simulations) the method could be useful in scenarios where efficiency is important and a Gaussian approximation is sufficient for uncertainty analysis. In a very high dimensional case, ADVI could become less efficient because of the increased size of the Gaussian covariance matrix. In that case one could use a mean-field approximation (setting model covariances to zero), or use a sparse covariance matrix to reduce computational cost since seismic velocity in any cell is often most strongly correlated with that in neighbouring cells.

SVGD can produce a good approximation to posterior pdfs. However, since it is based on a number of particles, the method is more computationally costly than ADVI. In this study we parallelized the computation of gradients to improve the efficiency, and for large datasets further improvements can be obtained by using random minibatches to perform the inversion \cite{liu2016stein}. Such a strategy can be applied to any variational inference method (e.g. also ADVI) since variational methods solve an optimization rather than a stochastic sampling problem. In comparison, this strategy cannot easily be used in McMC based methods since it may break the detailed balance requirement of McMC \cite{blei2017variational}. Though it has been shown that SVGD requires fewer particles than particle-based sampling methods (e.g., sequential Monte Carlo) in the sense that it reduces to finding the MAP model if only one particle is used, the optimal choice of the number of particles remains unclear, especially for very high dimensional spaces. In the case of very high dimensionality another possibility is to use normalizing flows -- a variational method based on a series of specific invertible transforms \cite{rezende2015variational}.

Monte Carlo and variational inference are different types of methods that solve the same problem. Monte Carlo simulates a set of Markov chains and uses samples of those chains to approximate the posterior pdf, while variational inference solves an optimization problem to find the closest pdf to the posterior within a given family of probability distributions. Monte Carlo methods provide guarantees that samples are asymptotically distributed according to the posterior pdf as the number of samples tends to infinity \cite{robert2013monte}, while the statistical properties of variational inference algorithms are still unknown \cite{blei2017variational}. It is possible to combine the two methods to capitalise on the merits of both. For example, the approximate posterior pdf from an efficient variational method (e.g. ADVI) can be used as a proposal distribution for Metropolis-Hastings \cite{de2001variational} to improve the efficiency of McMC, or McMC steps can be integrated to the variational approximation to improve the accuracy of variational methods \cite{salimans2015markov}.

We used a fixed regular grid of cells to parameterize the tomographic model in the variational methods, which might introduce overfitting of the data. For example, the mean velocity models in the synthetic tests show a slightly lower velocity loop between the low velocity anomaly and the receivers, and the uncertainties obtained from fixed-parameterization inversions are significantly higher than the results from rj-McMC. However, it is not easy to determine an optimal grid since this introduces a trade off between resolution of the model and overfitting of the data. Therefore, it might be necessary to use a more flexible parameterization, e.g., Voronoi cells \cite{bodin2009seismic,zhang20183} or wavelet parameterization \cite{fang2014wavelet, hawkins2015geophysical, zhang2015wavelet}. It may also be possible to apply a series of different parameterizations and select the best one using model selection theory \cite{walter1997identification, curtis1997reconditioning, arnold2018interrogation}.

In our experiments the results from rj-McMC are significantly different from the results obtained using variational methods or MH-McMC. This is essentially caused by different parameterizations. In ADVI, SVGD and MH-McMC we invert for a pixelated image, while in rj-McMC we invert for a distribution of parameters that represent locations and shapes of cells and their constant velocities, the pointwise spatial mean of which is visualized as an image. Therefore even though we visualized them in the same way, the results are essentially not directly comparable. Nevertheless, the comparison with rj-McMC is interesting because until now a quite different alternative probabilistic method was never used to estimate the posterior of images from the same realistic tomography problem. The results here demonstrate that the rj-McMC method as applied in most tomography papers gives significantly different solutions than we might previously have thought; specifically, it does not produce the posterior distribution of the pixelated image that is usually shown in scientific papers \cite[e.g.,][]{bodin2009seismic, galetti2015uncertainty, zulfakriza2014upper, crowder2019transdimensional}. Rather, it samples a probability distribution in a particular irregular and variably parametrized model space and results should be interpreted as such.

In this study we used a fixed data noise level in the variational methods. It has been shown that an improper noise level can introduce biases in tomographic results \cite{bodin2009seismic, zhang2019fully}, so in our example we used the noise level estimated by hierarchical McMC. It can also be estimated by a variety of other methods \cite{bensen20093, yao2009analysis, weaver2011precision, nicolson2012seismic, nicolson2014rayleigh}, and in future it might also be possible to include the noise parameters in variational methods in a hierarchical way.

In this study we applied variational inference methods to simple 2D tomography problems, but it is straightforward to apply the methods to any geophysical inverse problems whose gradients with respect to the model can be computed efficiently. For example, variational methods can be applied to 3D seismic tomography problems to provide efficient approximation, which generally demands enormous computational resources using McMC methods \cite{hawkins2015geophysical, zhang20183, zhang2019fully}. The methods also provide possibilities to perform Bayesian inference for full waveform inversion, which is generally very expensive for McMC \cite{ray2017low} and suffers from notorious multimodality in the likelihoods. SVGD provides a possible way to approximate these complex distributions given that theoretically it can approximate arbitrary distributions.
 
\section{Conclusion}
We introduced two variational inference methods to geophysical tomography -- automatic differential variational inference (ADVI) and Stein variational gradient descent (SVGD), and applied them to 2D seismic tomography problems using both synthetic and real data. Compared to the Markov chain Monte Carlo (McMC) method, ADVI provides an efficient but biased approximation to Bayesian posterior probability density functions. In contrast, SVGD is slightly slower than ADVI but produces a more accurate approximation. The real data example shows that ADVI and SVGD produce very similar mean velocity models, even though their uncertainty estimates are different because of a Gaussian approximation made implicitly within ADVI. The mean velocity models are very similar to those produced by reversible jump McMC (rj-McMC), except that the mean model from rj-McMC is smoother because of the much lower dimensionality of its parameter space. Variational methods thus can provide efficient approximate alternatives to McMC methods, and can be applied to many geophysical inverse problems.
 
\section*{Acknowledgments}
The authors would like to thank the Grane license partners Equinor ASA, Petoro AS, ExxonMobil E\&P Norway AS, and ConocoPhillips Skandinavia AS for allowing us to publish this work. The views and opinions expressed in this paper are those of the authors and are not necessarily shared by the license partners. The authors thank the Edinburgh Interferometry Project sponsors (Schlumberger, Equinor and Total) for supporting this research. This work used the Cirrus UK National Tier-2 HPC Service at EPCC (http://www.cirrus.ac.uk).

\bibliographystyle{plainnat}
\bibliography{variational}

\begin{thebibliography}{86}
\providecommand{\natexlab}[1]{#1}
\providecommand{\url}[1]{\texttt{#1}}
\expandafter\ifx\csname urlstyle\endcsname\relax
  \providecommand{\doi}[1]{doi: #1}\else
  \providecommand{\doi}{doi: \begingroup \urlstyle{rm}\Url}\fi

\bibitem[Aki and Lee(1976)]{aki1976determination}
Keiiti Aki and WHK Lee.
\newblock Determination of three-dimensional velocity anomalies under a seismic
  array using first {P} arrival times from local earthquakes: 1. a homogeneous
  initial model.
\newblock \emph{Journal of Geophysical research}, 81\penalty0 (23):\penalty0
  4381--4399, 1976.

\bibitem[Arnold and Curtis(2018)]{arnold2018interrogation}
Richard Arnold and Andrew Curtis.
\newblock Interrogation theory.
\newblock \emph{Geophysical Journal International}, 214\penalty0 (3):\penalty0
  1830--1846, 2018.

\bibitem[Bensen et~al.(2009)Bensen, Ritzwoller, and Yang]{bensen20093}
GD~Bensen, MH~Ritzwoller, and Y~Yang.
\newblock A 3-{D} shear velocity model of the crust and uppermost mantle
  beneath the {U}nited {S}tates from ambient seismic noise.
\newblock \emph{Geophysical Journal International}, 177\penalty0 (3):\penalty0
  1177--1196, 2009.

\bibitem[Bishop(2006)]{bishop2006pattern}
Christopher~M Bishop.
\newblock \emph{Pattern recognition and machine learning}.
\newblock springer, 2006.

\bibitem[Blei et~al.(2017)Blei, Kucukelbir, and McAuliffe]{blei2017variational}
David~M Blei, Alp Kucukelbir, and Jon~D McAuliffe.
\newblock Variational inference: A review for statisticians.
\newblock \emph{Journal of the American Statistical Association}, 112\penalty0
  (518):\penalty0 859--877, 2017.

\bibitem[Bodin and Sambridge(2009)]{bodin2009seismic}
Thomas Bodin and Malcolm Sambridge.
\newblock Seismic tomography with the reversible jump algorithm.
\newblock \emph{Geophysical Journal International}, 178\penalty0 (3):\penalty0
  1411--1436, 2009.

\bibitem[Bodin et~al.(2012)Bodin, Sambridge, Tkal{\v{c}}i{\'c}, Arroucau,
  Gallagher, and Rawlinson]{bodin2012transdimensional}
Thomas Bodin, Malcolm Sambridge, H~Tkal{\v{c}}i{\'c}, Pierre Arroucau, Kerry
  Gallagher, and Nicholas Rawlinson.
\newblock Transdimensional inversion of receiver functions and surface wave
  dispersion.
\newblock \emph{Journal of Geophysical Research: Solid Earth}, 117\penalty0
  (B2), 2012.

\bibitem[Burdick and Leki{\'c}(2017)]{burdick2017velocity}
Scott Burdick and Vedran Leki{\'c}.
\newblock Velocity variations and uncertainty from transdimensional {P}-wave
  tomography of {N}orth {A}merica.
\newblock \emph{Geophysical Journal International}, 209\penalty0 (2):\penalty0
  1337--1351, 2017.

\bibitem[Campillo and Paul(2003)]{campillo2003long}
Michel Campillo and Anne Paul.
\newblock Long-range correlations in the diffuse seismic coda.
\newblock \emph{Science}, 299\penalty0 (5606):\penalty0 547--549, 2003.

\bibitem[{\c{C}}{\i}nlar(2011)]{ccinlar2011probability}
Erhan {\c{C}}{\i}nlar.
\newblock \emph{Probability and stochastics}, volume 261.
\newblock Springer Science \& Business Media, 2011.

\bibitem[Crowder et~al.(2019)Crowder, Rawlinson, Pilia, Cornwell, and
  Reading]{crowder2019transdimensional}
E~Crowder, N~Rawlinson, S~Pilia, DG~Cornwell, and AM~Reading.
\newblock Transdimensional ambient noise tomography of {B}ass {S}trait,
  southeast {A}ustralia, reveals the sedimentary basin and deep crustal
  structure beneath a failed continental rift.
\newblock \emph{Geophysical Journal International}, 217\penalty0 (2):\penalty0
  970--987, 2019.

\bibitem[Curtis and Lomax(2001)]{curtis2001prior}
Andrew Curtis and Anthony Lomax.
\newblock Prior information, sampling distributions, and the curse of
  dimensionality.
\newblock \emph{Geophysics}, 66\penalty0 (2):\penalty0 372--378, 2001.

\bibitem[Curtis and Snieder(1997)]{curtis1997reconditioning}
Andrew Curtis and Roel Snieder.
\newblock Reconditioning inverse problems using the genetic algorithm and
  revised parameterization.
\newblock \emph{Geophysics}, 62\penalty0 (5):\penalty0 1524--1532, 1997.

\bibitem[Curtis and Snieder(2002)]{curtis2002probing}
Andrew Curtis and Roel Snieder.
\newblock Probing the earth's interior with seismic tomography.
\newblock \emph{International Geophysics Series}, 81\penalty0 (A):\penalty0
  861--874, 2002.

\bibitem[Curtis et~al.(2006)Curtis, Gerstoft, Sato, Snieder, and
  Wapenaar]{curtis2006seismic}
Andrew Curtis, Peter Gerstoft, Haruo Sato, Roel Snieder, and Kees Wapenaar.
\newblock Seismic interferometry -- turning noise into signal.
\newblock \emph{The Leading Edge}, 25\penalty0 (9):\penalty0 1082--1092, 2006.

\bibitem[De~Freitas et~al.(2001)De~Freitas, H{\o}jen-S{\o}rensen, Jordan, and
  Russell]{de2001variational}
Nando De~Freitas, Pedro H{\o}jen-S{\o}rensen, Michael~I Jordan, and Stuart
  Russell.
\newblock Variational {MCMC}.
\newblock In \emph{Proceedings of the Seventeenth conference on Uncertainty in
  artificial intelligence}, pages 120--127. Morgan Kaufmann Publishers Inc.,
  2001.

\bibitem[Devilee et~al.(1999)Devilee, Curtis, and
  Roy-Chowdhury]{devilee1999efficient}
RJR Devilee, A~Curtis, and K~Roy-Chowdhury.
\newblock An efficient, probabilistic neural network approach to solving
  inverse problems: Inverting surface wave velocities for {E}urasian crustal
  thickness.
\newblock \emph{Journal of Geophysical Research: Solid Earth}, 104\penalty0
  (B12):\penalty0 28841--28857, 1999.

\bibitem[Dziewonski and Woodhouse(1987)]{dziewonski1987global}
Adam~M Dziewonski and John~H Woodhouse.
\newblock Global images of the {E}arth's interior.
\newblock \emph{Science}, 236\penalty0 (4797):\penalty0 37--48, 1987.

\bibitem[Earp and Curtis(2019)]{earp2019probabilistic}
Stephanie Earp and Andrew Curtis.
\newblock Probabilistic neural-network based 2{D} travel time tomography.
\newblock \emph{arXiv preprint arXiv:1907.00541}, 2019.

\bibitem[Fang and Zhang(2014)]{fang2014wavelet}
Hongjian Fang and Haijiang Zhang.
\newblock Wavelet-based double-difference seismic tomography with sparsity
  regularization.
\newblock \emph{Geophysical Journal International}, 199\penalty0 (2):\penalty0
  944--955, 2014.

\bibitem[Galetti and Curtis(2018)]{galetti2018transdimensional}
E~Galetti and A~Curtis.
\newblock Transdimensional electrical resistivity tomography.
\newblock \emph{Journal of Geophysical Research: Solid Earth}, 123\penalty0
  (8):\penalty0 6347--6377, 2018.

\bibitem[Galetti et~al.(2015)Galetti, Curtis, Meles, and
  Baptie]{galetti2015uncertainty}
Erica Galetti, Andrew Curtis, Giovanni~Angelo Meles, and Brian Baptie.
\newblock Uncertainty loops in travel-time tomography from nonlinear wave
  physics.
\newblock \emph{Physical review letters}, 114\penalty0 (14):\penalty0 148501,
  2015.

\bibitem[Galetti et~al.(2017)Galetti, Curtis, Baptie, Jenkins, and
  Nicolson]{galetti2017transdimensional}
Erica Galetti, Andrew Curtis, Brian Baptie, David Jenkins, and Heather
  Nicolson.
\newblock Transdimensional love-wave tomography of the {B}ritish {I}sles and
  shear-velocity structure of the east {I}rish {S}ea {B}asin from ambient-noise
  interferometry.
\newblock \emph{Geophysical Journal International}, 208\penalty0 (1):\penalty0
  36--58, 2017.

\bibitem[Gorham and Mackey(2015)]{gorham2015measuring}
Jackson Gorham and Lester Mackey.
\newblock Measuring sample quality with {S}tein's method.
\newblock In \emph{Advances in Neural Information Processing Systems}, pages
  226--234, 2015.

\bibitem[Green(1995)]{green1995reversible}
Peter~J Green.
\newblock Reversible jump {M}arkov chain {M}onte {C}arlo computation and
  {B}yesian model determination.
\newblock \emph{Biometrika}, pages 711--732, 1995.

\bibitem[Green and Hastie(2009)]{green2009reversible}
Peter~J Green and David~I Hastie.
\newblock Reversible jump {MCMC}.
\newblock \emph{Genetics}, 155\penalty0 (3):\penalty0 1391--1403, 2009.

\bibitem[Gretton(2013)]{gretton2013introduction}
Arthur Gretton.
\newblock Introduction to {RKHS}, and some simple kernel algorithms.
\newblock 2013.

\bibitem[Hastings(1970)]{hastings1970monte}
W~Keith Hastings.
\newblock Monte {C}arlo sampling methods using {M}arkov chains and their
  applications.
\newblock \emph{Biometrika}, 57\penalty0 (1):\penalty0 97--109, 1970.

\bibitem[Hawkins and Sambridge(2015)]{hawkins2015geophysical}
Rhys Hawkins and Malcolm Sambridge.
\newblock Geophysical imaging using trans-dimensional trees.
\newblock \emph{Geophysical Journal International}, 203\penalty0 (2):\penalty0
  972--1000, 2015.

\bibitem[Hoffman and Blei(2015)]{hoffman2015structured}
Matthew~D Hoffman and David~M Blei.
\newblock Structured stochastic variational inference.
\newblock In \emph{Artificial Intelligence and Statistics}, 2015.

\bibitem[Iyer and Hirahara(1993)]{iyer1993seismic}
HM~Iyer and Kazuro Hirahara.
\newblock \emph{Seismic tomography: Theory and practice}.
\newblock Springer Science \& Business Media, 1993.

\bibitem[Karlin(2014)]{karlin2014first}
Samuel Karlin.
\newblock \emph{A first course in stochastic processes}.
\newblock Academic press, 2014.

\bibitem[K{\"a}ufl et~al.(2013)K{\"a}ufl, Valentine, O'Toole, and
  Trampert]{kaufl2013framework}
Paul K{\"a}ufl, Andrew~P Valentine, Thomas~B O'Toole, and Jeannot Trampert.
\newblock A framework for fast probabilistic centroid-moment-tensor
  determination -- inversion of regional static displacement measurements.
\newblock \emph{Geophysical Journal International}, 196\penalty0 (3):\penalty0
  1676--1693, 2013.

\bibitem[K{\"a}ufl et~al.(2015)K{\"a}ufl, Valentine, de~Wit, and
  Trampert]{kaufl2015robust}
Paul K{\"a}ufl, Andrew Valentine, Ralph de~Wit, and Jeannot Trampert.
\newblock Robust and fast probabilistic source parameter estimation from
  near-field displacement waveforms using pattern recognition.
\newblock \emph{Bulletin of the Seismological Society of America}, 105\penalty0
  (4):\penalty0 2299--2312, 2015.

\bibitem[Kingma and Welling(2013)]{kingma2013auto}
Diederik~P Kingma and Max Welling.
\newblock Auto-encoding variational {B}yes.
\newblock \emph{arXiv preprint arXiv:1312.6114}, 2013.

\bibitem[Kubrusly and Gravier(1973)]{kubrusly1973stochastic}
CS~Kubrusly and J~Gravier.
\newblock Stochastic approximation algorithms and applications.
\newblock In \emph{1973 IEEE conference on decision and control including the
  12th symposium on adaptive processes}, pages 763--766. IEEE, 1973.

\bibitem[Kucukelbir et~al.(2017)Kucukelbir, Tran, Ranganath, Gelman, and
  Blei]{kucukelbir2017automatic}
Alp Kucukelbir, Dustin Tran, Rajesh Ranganath, Andrew Gelman, and David~M Blei.
\newblock Automatic differentiation variational inference.
\newblock \emph{The Journal of Machine Learning Research}, 18\penalty0
  (1):\penalty0 430--474, 2017.

\bibitem[Kullback and Leibler(1951)]{kullback1951information}
Solomon Kullback and Richard~A Leibler.
\newblock On information and sufficiency.
\newblock \emph{The annals of mathematical statistics}, 22\penalty0
  (1):\penalty0 79--86, 1951.

\bibitem[Liu and Wang(2016)]{liu2016stein}
Qiang Liu and Dilin Wang.
\newblock Stein variational gradient descent: A general purpose {B}yesian
  inference algorithm.
\newblock In \emph{Advances In Neural Information Processing Systems}, pages
  2378--2386, 2016.

\bibitem[Liu et~al.(2016)Liu, Lee, and Jordan]{liu2016kernelized}
Qiang Liu, Jason Lee, and Michael Jordan.
\newblock A kernelized {S}tein discrepancy for goodness-of-fit tests.
\newblock In \emph{International Conference on Machine Learning}, pages
  276--284, 2016.

\bibitem[Malinverno(2002)]{malinverno2002parsimonious}
Alberto Malinverno.
\newblock Parsimonious {B}yesian {M}arkov chain {M}onte {C}arlo inversion in a
  nonlinear geophysical problem.
\newblock \emph{Geophysical Journal International}, 151\penalty0 (3):\penalty0
  675--688, 2002.

\bibitem[Malinverno and Briggs(2004)]{malinverno2004expanded}
Alberto Malinverno and Victoria~A Briggs.
\newblock Expanded uncertainty quantification in inverse problems: Hierarchical
  {B}yes and empirical {B}yes.
\newblock \emph{Geophysics}, 69\penalty0 (4):\penalty0 1005--1016, 2004.

\bibitem[Malinverno et~al.(2000)Malinverno, Leaney,
  et~al.]{malinverno2000monte}
Alberto Malinverno, Scott Leaney, et~al.
\newblock A {M}onte {C}arlo method to quantify uncertainty in the inversion of
  zero-offset {VSP} data.
\newblock In \emph{2000 SEG Annual Meeting}. Society of Exploration
  Geophysicists, 2000.

\bibitem[Marzouk et~al.(2016)Marzouk, Moselhy, Parno, and
  Spantini]{marzouk2016introduction}
Youssef Marzouk, Tarek Moselhy, Matthew Parno, and Alessio Spantini.
\newblock An introduction to sampling via measure transport.
\newblock \emph{arXiv preprint arXiv:1602.05023}, 2016.

\bibitem[Meier et~al.(2007{\natexlab{a}})Meier, Curtis, and
  Trampert]{meier2007bglobal}
U~Meier, A~Curtis, and J~Trampert.
\newblock A global crustal model constrained by nonlinearised inversion of
  fundamental mode surface waves.
\newblock \emph{Geophysical Research Letters}, 34:\penalty0 L16304,
  2007{\natexlab{a}}.

\bibitem[Meier et~al.(2007{\natexlab{b}})Meier, Curtis, and
  Trampert]{meier2007aglobal}
Ueli Meier, Andrew Curtis, and Jeannot Trampert.
\newblock Global crustal thickness from neural network inversion of surface
  wave data.
\newblock \emph{Geophysical Journal International}, 169\penalty0 (2):\penalty0
  706--722, 2007{\natexlab{b}}.

\bibitem[Metropolis and Ulam(1949)]{metropolis1949monte}
Nicholas Metropolis and Stanislaw Ulam.
\newblock The {M}onte {C}arlo method.
\newblock \emph{Journal of the American statistical association}, 44\penalty0
  (247):\penalty0 335--341, 1949.

\bibitem[Mosegaard and Tarantola(1995)]{mosegaard1995monte}
Klaus Mosegaard and Albert Tarantola.
\newblock Monte {C}arlo sampling of solutions to inverse problems.
\newblock \emph{Journal of Geophysical Research: Solid Earth}, 100\penalty0
  (B7):\penalty0 12431--12447, 1995.

\bibitem[Nawaz and Curtis(2019)]{nawaz2019rapid}
MA~Nawaz and A~Curtis.
\newblock Rapid discriminative variational {B}yesian inversion of geophysical
  data for the spatial distribution of geological properties.
\newblock \emph{Journal of Geophysical Research: Solid Earth}, 2019.

\bibitem[Nawaz and Curtis(2018)]{nawaz2018variational}
Muhammad~Atif Nawaz and Andrew Curtis.
\newblock Variational {B}ayesian inversion ({VBI}) of quasi-localized seismic
  attributes for the spatial distribution of geological facies.
\newblock \emph{Geophysical Journal International}, 214\penalty0 (2):\penalty0
  845--875, 2018.

\bibitem[Nicolson et~al.(2012)Nicolson, Curtis, Baptie, and
  Galetti]{nicolson2012seismic}
Heather Nicolson, Andrew Curtis, Brian Baptie, and Erica Galetti.
\newblock Seismic interferometry and ambient noise tomography in the {B}ritish
  {I}sles.
\newblock \emph{Proceedings of the Geologists' Association}, 123\penalty0
  (1):\penalty0 74--86, 2012.

\bibitem[Nicolson et~al.(2014)Nicolson, Curtis, and
  Baptie]{nicolson2014rayleigh}
Heather Nicolson, Andrew Curtis, and Brian Baptie.
\newblock Rayleigh wave tomography of the {B}ritish {I}sles from ambient
  seismic noise.
\newblock \emph{Geophysical Journal International}, 198\penalty0 (2):\penalty0
  637--655, 2014.

\bibitem[Piana~Agostinetti et~al.(2015)Piana~Agostinetti, Giacomuzzi, and
  Malinverno]{piana2015local}
Nicola Piana~Agostinetti, Genny Giacomuzzi, and Alberto Malinverno.
\newblock Local three-dimensional earthquake tomography by trans-dimensional
  {M}onte {C}arlo sampling.
\newblock \emph{Geophysical Journal International}, 201\penalty0 (3):\penalty0
  1598--1617, 2015.

\bibitem[Ranganath et~al.(2014)Ranganath, Gerrish, and
  Blei]{ranganath2014black}
Rajesh Ranganath, Sean Gerrish, and David Blei.
\newblock Black box variational inference.
\newblock In \emph{Artificial Intelligence and Statistics}, pages 814--822,
  2014.

\bibitem[Ranganath et~al.(2016)Ranganath, Tran, and
  Blei]{ranganath2016hierarchical}
Rajesh Ranganath, Dustin Tran, and David Blei.
\newblock Hierarchical variational models.
\newblock In \emph{International Conference on Machine Learning}, pages
  324--333, 2016.

\bibitem[Rawlinson and Sambridge(2004)]{rawlinson2004multiple}
Nick Rawlinson and Malcolm Sambridge.
\newblock Multiple reflection and transmission phases in complex layered media
  using a multistage fast marching method.
\newblock \emph{Geophysics}, 69\penalty0 (5):\penalty0 1338--1350, 2004.

\bibitem[Ray et~al.(2013)Ray, Alumbaugh, Hoversten, and Key]{ray2013robust}
Anandaroop Ray, David~L Alumbaugh, G~Michael Hoversten, and Kerry Key.
\newblock Robust and accelerated {B}yesian inversion of marine
  controlled-source electromagnetic data using parallel tempering.
\newblock \emph{Geophysics}, 78\penalty0 (6):\penalty0 E271--E280, 2013.

\bibitem[Ray et~al.(2017)Ray, Kaplan, Washbourne, and Albertin]{ray2017low}
Anandaroop Ray, Sam Kaplan, John Washbourne, and Uwe Albertin.
\newblock Low frequency full waveform seismic inversion within a tree based
  {B}yesian framework.
\newblock \emph{Geophysical Journal International}, 212\penalty0 (1):\penalty0
  522--542, 2017.

\bibitem[Rezende and Mohamed(2015)]{rezende2015variational}
Danilo~Jimenez Rezende and Shakir Mohamed.
\newblock Variational inference with normalizing flows.
\newblock \emph{arXiv preprint arXiv:1505.05770}, 2015.

\bibitem[Robbins and Monro(1951)]{robbins1951stochastic}
Herbert Robbins and Sutton Monro.
\newblock A stochastic approximation method.
\newblock \emph{The annals of mathematical statistics}, pages 400--407, 1951.

\bibitem[Robert and Casella(2013)]{robert2013monte}
Christian Robert and George Casella.
\newblock \emph{Monte {C}arlo statistical methods}.
\newblock Springer Science \& Business Media, 2013.

\bibitem[R{\"o}th and Tarantola(1994)]{roth1994neural}
Gunter R{\"o}th and Albert Tarantola.
\newblock Neural networks and inversion of seismic data.
\newblock \emph{Journal of Geophysical Research: Solid Earth}, 99\penalty0
  (B4):\penalty0 6753--6768, 1994.

\bibitem[Salimans et~al.(2015)Salimans, Kingma, and
  Welling]{salimans2015markov}
Tim Salimans, Diederik Kingma, and Max Welling.
\newblock Markov chain {M}onte {C}arlo and variational inference: Bridging the
  gap.
\newblock In \emph{International Conference on Machine Learning}, pages
  1218--1226, 2015.

\bibitem[Sambridge(1999)]{sambridge1999geophysical}
Malcolm Sambridge.
\newblock Geophysical inversion with a neighbourhood algorithm -- i. searching
  a parameter space.
\newblock \emph{Geophysical journal international}, 138\penalty0 (2):\penalty0
  479--494, 1999.

\bibitem[Saul and Jordan(1996)]{saul1996exploiting}
Lawrence~K Saul and Michael~I Jordan.
\newblock Exploiting tractable substructures in intractable networks.
\newblock In \emph{Advances in neural information processing systems}, pages
  486--492, 1996.

\bibitem[Shahraeeni and Curtis(2011)]{shahraeeni2011fast}
Mohammad~S Shahraeeni and Andrew Curtis.
\newblock Fast probabilistic nonlinear petrophysical inversion.
\newblock \emph{Geophysics}, 76\penalty0 (2):\penalty0 E45--E58, 2011.

\bibitem[Shahraeeni et~al.(2012)Shahraeeni, Curtis, and
  Chao]{shahraeeni2012fast}
Mohammad~S Shahraeeni, Andrew Curtis, and Gabriel Chao.
\newblock Fast probabilistic petrophysical mapping of reservoirs from 3{D}
  seismic data.
\newblock \emph{Geophysics}, 77\penalty0 (3):\penalty0 O1--O19, 2012.

\bibitem[Shapiro et~al.(2005)Shapiro, Campillo, Stehly, and
  Ritzwoller]{shapiro2005high}
Nikolai~M Shapiro, Michel Campillo, Laurent Stehly, and Michael~H Ritzwoller.
\newblock High-resolution surface-wave tomography from ambient seismic noise.
\newblock \emph{Science}, 307\penalty0 (5715):\penalty0 1615--1618, 2005.

\bibitem[Shen et~al.(2012)Shen, Ritzwoller, Schulte-Pelkum, and
  Lin]{shen2012joint}
Weisen Shen, Michael~H Ritzwoller, Vera Schulte-Pelkum, and Fan-Chi Lin.
\newblock Joint inversion of surface wave dispersion and receiver functions: a
  {B}yesian {M}onte-{C}arlo approach.
\newblock \emph{Geophysical Journal International}, 192\penalty0 (2):\penalty0
  807--836, 2012.

\bibitem[Shen et~al.(2013)Shen, Ritzwoller, and Schulte-Pelkum]{shen20133}
Weisen Shen, Michael~H Ritzwoller, and Vera Schulte-Pelkum.
\newblock A 3-{D} model of the crust and uppermost mantle beneath the central
  and western {US} by joint inversion of receiver functions and surface wave
  dispersion.
\newblock \emph{Journal of Geophysical Research: Solid Earth}, 118\penalty0
  (1):\penalty0 262--276, 2013.

\bibitem[Sivia(1996)]{sivia1996data}
DS~Sivia.
\newblock Data analysis: A {B}yesian tutorial (oxford science publications).
\newblock 1996.

\bibitem[Smith(2013)]{smith2013sequential}
Adrian Smith.
\newblock \emph{Sequential {M}onte {C}arlo methods in practice}.
\newblock Springer Science \& Business Media, 2013.

\bibitem[Stein et~al.(1972)]{stein1972bound}
Charles Stein et~al.
\newblock A bound for the error in the normal approximation to the distribution
  of a sum of dependent random variables.
\newblock In \emph{Proceedings of the Sixth Berkeley Symposium on Mathematical
  Statistics and Probability, Volume 2: Probability Theory}. The Regents of the
  University of California, 1972.

\bibitem[Tarantola(2005)]{tarantola2005inverse}
Albert Tarantola.
\newblock \emph{Inverse problem theory and methods for model parameter
  estimation}, volume~89.
\newblock SIAM, 2005.

\bibitem[Team et~al.(2016)]{stan2016stan}
Stan~Development Team et~al.
\newblock Stan modeling language users guide and reference manual.
\newblock \emph{Technical report}, 2016.

\bibitem[Tran et~al.(2015)Tran, Ranganath, and Blei]{tran2015variational}
Dustin Tran, Rajesh Ranganath, and David~M Blei.
\newblock The variational {G}aussian process.
\newblock \emph{arXiv preprint arXiv:1511.06499}, 2015.

\bibitem[Walter and Pronzato(1997)]{walter1997identification}
Eric Walter and Luc Pronzato.
\newblock \emph{Identification of parametric models from experimental data}.
\newblock Springer Verlag, 1997.

\bibitem[Weaver et~al.(2011)Weaver, Hadziioannou, Larose, and
  Campillo]{weaver2011precision}
Richard~L Weaver, C{\'e}line Hadziioannou, Eric Larose, and Michel Campillo.
\newblock On the precision of noise correlation interferometry.
\newblock \emph{Geophysical Journal International}, 185\penalty0 (3):\penalty0
  1384--1392, 2011.

\bibitem[Yao and Van Der~Hilst(2009)]{yao2009analysis}
Huajian Yao and Robert~D Van Der~Hilst.
\newblock Analysis of ambient noise energy distribution and phase velocity bias
  in ambient noise tomography, with application to {SE} tibet.
\newblock \emph{Geophysical Journal International}, 179\penalty0 (2):\penalty0
  1113--1132, 2009.

\bibitem[Young et~al.(2013)Young, Rawlinson, and
  Bodin]{young2013transdimensional}
Mallory~K Young, Nicholas Rawlinson, and Thomas Bodin.
\newblock Transdimensional inversion of ambient seismic noise for 3{D} shear
  velocity structure of the {T}asmanian crust.
\newblock \emph{Geophysics}, 78\penalty0 (3):\penalty0 WB49--WB62, 2013.

\bibitem[Zhang et~al.(2019)Zhang, Hansteen, and Curtis]{zhang2019fully}
X~Zhang, F~Hansteen, and A~Curtis.
\newblock Fully 3{D} {M}onte {C}arlo ambient noise tomography over {G}rane
  field.
\newblock In \emph{81st EAGE Conference and Exhibition 2019}, 2019.

\bibitem[Zhang and Zhang(2015)]{zhang2015wavelet}
Xin Zhang and Haijiang Zhang.
\newblock Wavelet-based time-dependent travel time tomography method and its
  application in imaging the {E}tna volcano in {I}taly.
\newblock \emph{Journal of Geophysical Research: Solid Earth}, 120\penalty0
  (10):\penalty0 7068--7084, 2015.

\bibitem[Zhang et~al.(2018)Zhang, Curtis, Galetti, and de~Ridder]{zhang20183}
Xin Zhang, Andrew Curtis, Erica Galetti, and Sjoerd de~Ridder.
\newblock 3-{D} {M}onte {C}arlo surface wave tomography.
\newblock \emph{Geophysical Journal International}, 215\penalty0 (3):\penalty0
  1644--1658, 2018.

\bibitem[Zhdanov(2002)]{zhdanov2002geophysical}
Michael~S Zhdanov.
\newblock \emph{Geophysical inverse theory and regularization problems},
  volume~36.
\newblock Elsevier, 2002.

\bibitem[Zheng et~al.(2017)Zheng, Saygin, Cummins, Ge, Min, Cipta, and
  Yang]{zheng2017transdimensional}
DingChang Zheng, Erdinc Saygin, Phil Cummins, Zengxi Ge, Zhaoxu Min, Athanasius
  Cipta, and Runhai Yang.
\newblock Transdimensional {B}yesian seismic ambient noise tomography across
  {SE} tibet.
\newblock \emph{Journal of Asian Earth Sciences}, 134:\penalty0 86--93, 2017.

\bibitem[Zulfakriza et~al.(2014)Zulfakriza, Saygin, Cummins, Widiyantoro,
  Nugraha, L{\"u}hr, and Bodin]{zulfakriza2014upper}
Z~Zulfakriza, Erdinc Saygin, PR~Cummins, Sri Widiyantoro, Andri~Dian Nugraha,
  B-G L{\"u}hr, and T~Bodin.
\newblock Upper crustal structure of central {J}ava, {I}ndonesia, from
  transdimensional seismic ambient noise tomography.
\newblock \emph{Geophysical Journal International}, 197\penalty0 (1):\penalty0
  630--635, 2014.

\end{thebibliography}

\appendix
\section{The entropy of a Gaussian distribution}

The entropy $\mathrm{H} \left[ q(\boldsymbol{\theta};\boldsymbol{\phi}) \right]$ of a Gaussian distribution $\mathcal{N}(\boldsymbol{\theta}|\boldsymbol{\mu},\mathbf{L}\mathbf{L}^{T})$ is:
\begin{align*}
\mathrm{H} \left[ q(\boldsymbol{\theta};\boldsymbol{\phi}) \right] &= -\mathrm{E}_{q}[\mathrm{log}q(\boldsymbol{\theta})] \\
&= - \int \mathcal{N}(\boldsymbol{\theta}|\boldsymbol{\mu},\mathbf{L}\mathbf{L}^{T}) \mathrm{log}\mathcal{N}(\boldsymbol{\theta}|\boldsymbol{\mu},\mathbf{L}\mathbf{L}^{T})d\boldsymbol{\theta} \\
&= \frac{k}{2} + \frac{k}{2} \mathrm{log}(2\pi) + \frac{1}{2} \mathrm{log} |det( \mathbf{LL}^{T})|
\end{align*}
where $k$ is the dimension of vector $\boldsymbol{\theta}$. The gradients with respect to $\boldsymbol{\mu}$ and $\mathbf{L}$ can be easily calculated (see Appendix B).

\section{Gradients of the ELBO in ADVI}
We first describe the dominated convergence theorem (DCT) \cite{ccinlar2011probability}:

\noindent
\textbf{Theorem} Assume $X \in \mathcal{X}$ is a random variable and $f: \mathbb{R} \times \mathcal{X} \to \mathbb{R}$ is a function such that $f(t,X)$ is integrable for all $t$ and $\frac{\partial f(t,X)}{\partial t}$ exists for each $t$. Assume that there is a random variable $Z$ such that $|\frac{\partial f(t,X)}{\partial t}| \le Z$ for all $t$ and $\mathrm{E}(Z)<\infty$. Then
\begin{equation*}
   \frac{\partial}{\partial t} \mathrm{E}(f(t,X)) = \mathrm{E}(\frac{\partial}{\partial t}f(t,X))
\end{equation*}
The proof of this theorem is given in \citet{ccinlar2011probability}.

We then calculate the gradients in equation (\ref{eq:gradient_mu}) and (\ref{eq:gradient_sigma}) based on \citet{kucukelbir2017automatic}. The ELBO $\mathcal{L}$ is:
\begin{equation*}
 \mathcal{L} = \mathrm{E}_{ \mathcal{N} (\boldsymbol{\eta} | \boldsymbol{0},\mathbf{I} ) } \left[ \mathrm{log} p\big( T^{-1} \left( R_{\boldsymbol{\phi}}^{-1} (\boldsymbol{\eta}) \right), \mathbf{d}_{obs} \big) + \mathrm{log} |det\mathbf{J}_{T^{-1}} \left(R_{\boldsymbol{\phi}}^{-1} (\boldsymbol{\eta} )\right) | \right] + \mathrm{H} \left[ q(\boldsymbol{\theta};\boldsymbol{\phi}) \right] 
\end{equation*} 
where $\mathrm{H} \left[ q(\boldsymbol{\theta};\boldsymbol{\phi}) \right] = \mathrm{E}_{q} \left[ \mathrm{log}q(\boldsymbol{\theta} \right]$ is the entropy of distribution $q$. Assume $\frac{\partial}{\partial \boldsymbol{\phi}} \mathrm{log}p$ is bounded where $\boldsymbol{\phi}$ represents variational parameters $\boldsymbol{\mu}$ and $\mathbf{L}$, then the gradients can be computed by exchanging the derivative and the expectation using the dominated convergence theorem (DCT) and applying the chain rule:
\begin{align*}
\begin{split}
 \nabla_{\boldsymbol{\mu}}\mathcal{L} 
 &= 
 \nabla_{ \boldsymbol{\mu} } 
 \left\{ 
  \mathrm{E}_{ \mathcal{N} (\boldsymbol{\eta} | \boldsymbol{0},\mathbf{I} ) } 
  \left[ 
    \mathrm{log} p\big( T^{-1} \left( R_{\boldsymbol{\phi}}^{-1} (\boldsymbol{\eta}) \right), \mathbf{d}_{obs} \big) 
    + 
    \mathrm{log} |det\mathbf{J}_{T^{-1}} \left(R_{\boldsymbol{\phi}}^{-1} (\boldsymbol{\eta} )\right) | 
  \right] 
  + 
  \mathrm{H} \left[ q(\boldsymbol{\theta};\boldsymbol{\phi}) \right] 
 \right\} \\ 
 \intertext{Applying the DCT and since H does not depend on $\boldsymbol{\mu}$,} 
 \nabla_{\boldsymbol{\mu}}\mathcal{L} &= 
 \mathrm{E}_{ \mathcal{N} (\boldsymbol{\eta} | \boldsymbol{0},\mathbf{I} ) } 
   \left[ 
     \nabla_{ \boldsymbol{\mu} } \left\{  
           \mathrm{log} p\big( T^{-1} \left( R_{\boldsymbol{\phi}}^{-1} (\boldsymbol{\eta}) \right), \mathbf{d}_{obs} \big) \right\} 
     + 
     \nabla_{ \boldsymbol{\mu} } \left( 
           \mathrm{log} |det\mathbf{J}_{T^{-1}} \left(R_{\boldsymbol{\phi}}^{-1} (\boldsymbol{\eta} ) \right) | \right) 
    \right] \\
 \intertext{Applying the chain rule,}
\nabla_{\boldsymbol{\mu}}\mathcal{L}  &=
 \mathrm{E}_{\mathcal{N}(\boldsymbol{\eta}|\boldsymbol{0},\mathbf{I})} 
 \left[ 
 \nabla_{\mathbf{m}} \mathrm{log}p(\mathbf{m},\mathbf{d}_{obs}) 
 \nabla_{\boldsymbol{\theta}}T^{-1}(\boldsymbol{\theta}) 
 \nabla_{\boldsymbol{\mu}} R_{\boldsymbol{\phi}}^{-1} (\boldsymbol{\eta} ) 
 + 
 \nabla_{\boldsymbol{\theta}} \mathrm{log}|det\mathbf{J}_{T^{-1}}(\boldsymbol{\theta})| 
 \nabla_{\boldsymbol{\mu}} R_{\boldsymbol{\phi}}^{-1} (\boldsymbol{\eta})
 \right] \\
 &=
 \mathrm{E}_{\mathcal{N}(\boldsymbol{\eta}|\boldsymbol{0},\mathbf{I})} 
 \left[ 
 \nabla_{\mathbf{m}}\mathrm{log}p(\mathbf{m},\mathbf{d}_{obs}) 
 \nabla_{\boldsymbol{\theta}}T^{-1}(\boldsymbol{\theta})
 +
 \nabla_{\boldsymbol{\theta}}\mathrm{log}|det\mathbf{J}_{T^{-1}}(\boldsymbol{\theta})| 
 \right]
\end{split}
 \end{align*}
The gradient with respect to $\mathbf{L}$ can be obtained similarly,
\begin{align*}
\begin{split}
 \nabla_{\mathbf{L}}\mathcal{L}
  &= 
  \nabla_{ \mathbf{L} } 
  \Big\{ 
  \mathrm{E}_{ \mathcal{N} (\boldsymbol{\eta} | \boldsymbol{0},\mathbf{I} ) } 
  \left[ 
    \mathrm{log} p\left( T^{-1} \left( R_{\boldsymbol{\phi}}^{-1} (\boldsymbol{\eta}) \right), \mathbf{d}_{obs} \right) 
     + 
    \mathrm{log} |det\mathbf{J}_{T^{-1}} \left(R_{\boldsymbol{\phi}}^{-1} (\boldsymbol{\eta} )\right) | 
  \right] \\ 
  & \qquad +
  \frac{k}{2} + \frac{k}{2} \mathrm{log}(2\pi) + \frac{1}{2} \mathrm{log} |det( \mathbf{LL}^{T})| 
  \Big\}
\end{split}
\\
\intertext{Applying the DCT}
\begin{split}
\nabla_{\mathbf{L}}\mathcal{L} &= 
 \mathrm{E}_{ \mathcal{N} (\boldsymbol{\eta} | \boldsymbol{0},\mathbf{I} ) } 
   \big[ 
     \nabla_{ \mathbf{L} } \left\{  
           \mathrm{log} p\left( T^{-1} \left( R_{\boldsymbol{\phi}}^{-1} (\boldsymbol{\eta}) \right), \mathbf{d}_{obs} \right) \right\} 
     + 
     \nabla_{ \mathbf{L} } \left( 
           \mathrm{log} |det\mathbf{J}_{T^{-1}} \left(R_{\boldsymbol{\phi}}^{-1} (\boldsymbol{\eta} ) \right) | \right) \big] \\
    &\qquad + \nabla_{ \mathbf{L} } \frac{1}{2} \mathrm{log} |det( \mathbf{LL}^{T})|    
\end{split}
 \\
\intertext{and applying the chain rule we obtain} 
\nabla_{\mathbf{L}}\mathcal{L} &=
 \mathrm{E}_{\mathcal{N}(\boldsymbol{\eta}|\boldsymbol{0},\mathbf{I})} 
 \left[ 
 \nabla_{\mathbf{m}} \mathrm{log}p(\mathbf{m},\mathbf{d}_{obs}) 
 \nabla_{\boldsymbol{\theta}}T^{-1}(\boldsymbol{\theta}) 
 \nabla_{\mathbf{L} } R_{\boldsymbol{\phi}}^{-1} (\boldsymbol{\eta} ) 
 + 
 \nabla_{\boldsymbol{\theta}} \mathrm{log}|det\mathbf{J}_{T^{-1}}(\boldsymbol{\theta})| 
 \nabla_{\mathbf{L}} R_{\boldsymbol{\phi}}^{-1} (\boldsymbol{\eta})
 \right] 
 + (\mathbf{L}^{-1})^{T} \\
 &= 
 \mathrm{E}_{\mathcal{N}(\boldsymbol{\eta}|\boldsymbol{0},\mathbf{I})}
 \left[ 
 \nabla_{\mathbf{m}}\mathrm{log} p(\mathbf{m},\mathbf{d}_{obs}) 
 \nabla_{\boldsymbol{\theta}}T^{-1}(\boldsymbol{\theta})
 +
 \nabla_{\boldsymbol{\theta}}\mathrm{log}|det\mathbf{J}_{T^{-1}}(\boldsymbol{\theta})|\boldsymbol{\eta}^{T} 
 \right] 
 + 
 (\mathbf{L}^{-1})^{T}
\end{align*}

\section{Gradients of KL-divergence in SVGD}
We calculate the gradient in equation (\ref{eq:stein_gradient}) following \citet{liu2016stein}. Denote $T^{-1}$ as the inverse transform of $T$. Then by changing the variable,
\begin{equation*}
  \mathrm{KL}[q_{T}||p] = \mathrm{KL}[q||p_{T^{-1}}]
\end{equation*}
and hence
\begin{align*}
\nabla_{\epsilon} \mathrm{KL}[q_{T}||p] \, |_{\epsilon=0}
&= \nabla_{\epsilon} \mathrm{KL}[q||p_{T^{-1}}] \, |_{\epsilon=0} \\
&= \nabla_{\epsilon} \left[ \mathrm{E}_{q} \mathrm{log}q(\mathbf{m}) - \mathrm{E}_{q} \mathrm{log}p_{T^{-1}} (\mathbf{m})  \right] \\
\intertext{and since $q(\mathbf{m})$ does not depend on $\epsilon$}
\nabla_{\epsilon} \mathrm{KL}[q_{T}||p] \, |_{\epsilon=0} &= - \mathrm{E}_{q} \left[ \nabla_{\epsilon} \mathrm{log} p_{T^{-1}} (\mathbf{m})  \right]
\end{align*}
where $p_{T^{-1}}(\mathbf{m}) = p(T(\mathbf{m})) \cdot |\mathrm{det} \left( \nabla_{\mathbf{m}} T(\mathbf{m}) \right)|$. Therefore
\begin{equation*}
\nabla_{\epsilon} \mathrm{log} p_{T^{-1}} (\mathbf{m}) 
= \left( \nabla_{\mathbf{m}} \mathrm{log} \left(p(\mathbf{m}) \right) \right)^{\mathrm{T}} \nabla_{\epsilon} T(\mathbf{m})
 + trace\left( \left( \nabla_{\mathbf{m}} T(\mathbf{m}) \right)^{-1} \cdot \nabla_{\epsilon}\nabla_{\mathbf{m}} T(\mathbf{m}) \right) 
\end{equation*}
where $T(\mathbf{m}) = \mathbf{m} + \epsilon\boldsymbol{\phi}(\mathbf{m})$, $\nabla_{\epsilon} T(\mathbf{m}=\boldsymbol{\phi}(\mathbf{m}) $ and $\nabla_{\mathbf{m}} T(\mathbf{m})|_{\epsilon=0} = \mathbf{I} $, and so
\begin{align*}
\nabla_{\epsilon} \mathrm{KL}[q_{T}||p] \, |_{\epsilon=0}
 &= - \mathrm{E}_{q} \left[ \left( \nabla_{\mathbf{m}} \mathrm{log} \left(p(\mathbf{m}) \right) \right)^{\mathrm{T}} \boldsymbol{\phi}(\mathbf{m})
    + trace\left( \nabla_{\mathbf{m}} \boldsymbol{\phi}(\mathbf{m}) \right) \right] \\
 &= - \mathrm{E}_{q} \left[ trace\left( \nabla_{\mathbf{m}} \mathrm{log} \left(p(\mathbf{m}) \right) \boldsymbol{\phi}(\mathbf{m})^{\mathbf{T}} \right)
    + trace\left( \nabla_{\mathbf{m}} \boldsymbol{\phi}(\mathbf{m}) \right) \right] \\
 &= - \mathrm{E}_{q} \left[ trace \left( \mathcal{A}_{p} \boldsymbol{\phi} (\mathbf{m}) \right) \right] 
\end{align*} 
where $\mathcal{A}_{p} \boldsymbol{\phi}(\mathbf{m}) = \nabla_{\mathbf{m}} \mathrm{log} p(\mathbf{m}) \boldsymbol{\phi} (\mathbf{m})^{T} + \nabla_{ \mathbf{m} } \boldsymbol{\phi} ( \mathbf{m} )$ is the Stein operator.

\label{lastpage}

\end{document}